\documentclass[5p,twocolumn,authoryear]{elsarticle}

\usepackage{hyperref}
\usepackage[utf8]{inputenc}
\usepackage{amsmath}
\usepackage{amsfonts}
\usepackage{amssymb}
\usepackage{graphicx}
\usepackage{color}
\usepackage{url}
\usepackage[T1]{fontenc}
\usepackage{subcaption}
\usepackage{algpseudocode}
\usepackage{algorithm}
\usepackage{float}
\usepackage{pdflscape}

\algnewcommand\algorithmicforeach{\textbf{for each}}
\algdef{S}[FOR]{ForEach}[1]{\algorithmicforeach\ #1\ \algorithmicdo}

\usepackage{balance}
\usepackage{listings}
\usepackage{fancybox}
\usepackage{multirow}
\usepackage{cleveref}
\usepackage{flafter,placeins}

\newcommand{\comment}[1]{\textcolor{black}{{#1}}}

\usepackage{xspace}

\journal{Journal of Sustainable Computing: Informatics and Systems}

\begin{document}
\FloatBarrier

\begin{frontmatter}

\title{An App Performance Optimization Advisor for Mobile Device App Marketplaces}

\author{Rub\'{e}n Saborido\corref{cor1}} 
\ead{ruben.saborido-infantes@polymtl.ca}
\cortext[cor1]{Corresponding author}

\author{Foutse Khomh}
\ead{foutse.khomh@polymtl.ca}
\address{DGIGL, Polytechnique Montr\'{e}al, Quebec (Canada)}

\author{Abram Hindle}
\ead{abram.hindle@ualberta.ca}
\address{University of Alberta, Alberta (Canada)}

\author{Enrique Alba}
\ead{eat@lcc.uma.es}
\address{University of Málaga, Málaga (Spain)}

\begin{abstract}
On mobile phones, users and developers use apps official marketplaces serving as repositories of apps. The Google Play Store and Apple Store are the official marketplaces of Android and Apple products which offer more than a million apps. Although both repositories offer description of apps, information concerning performance is not available. Due to the constrained hardware of mobile devices, users and developers have to meticulously manage the resources available and they should be given access to performance information about apps. Even if this information was available, the selection of apps would still depend on user preferences and it would require a huge cognitive effort to make optimal decisions. Considering this fact we propose APOA, a recommendation system which can be implemented in any marketplace for helping users and developers to compare apps in terms of performance.

APOA uses as input metric values of apps and a set of metrics to optimize. It solves an optimization problem and it generates optimal sets of apps  for different user's context. We show how APOA works over an Android case study. Out of 140 apps, we define typical usage scenarios and we collect measurements of power, CPU, memory, and network usages to demonstrate the benefit of using APOA.

\end{abstract}

\end{frontmatter}

\section{Introduction}
\label{sec:introduction}


Typical users of mobile devices purchase and download apps
using platform dependent repositories/marketplaces of apps, colloquially
referred to as app-stores.
Android is an open-source operating system for mobile devices. It is used by more than 1.4 billion users for a global market share of 53\%\footnote{http://expandedramblings.com/index.php/android-statistics/}. 
iOS is a mobile operating system created and developed by Apple for its hardware. 
It is the second most popular mobile operating system globally after Android\footnote{http://www.kpcb.com/internet-trends}. 
Both platforms offer apps belonging to different categories through their marketplaces; the Google Play Store \footnote{https://play.google.com/store/apps} 
and the Apple Store\footnote{https://itunes.apple.com/ca/genre/ios/id36?mt=8} \comment{apps}
, for Android and iOS, respectively. For each app in any of these marketplaces, customer ratings are provided as a quality metric. The rating is a number between one and five, which is calculated as the weighted average of user ratings in the marketplace. Mobile device users compare and select apps from marketplaces based on their rating and number of downloads~(\cite{gomes_empirical_2016}). This fact results in users choosing apps that other users choose (popularity), even if the apps are less efficient than other apps offering similar functionalities (\cite{saborido_optimizing_2016}).

Selecting a cost-effective app is a challenging task, because of the large number of apps that offer similar functionalities, and the lack of information about the performance of the apps. 
For example, there are a very large number of browsers, cameras, and music players available but the decision on which app to use lies more on subjective requirements, such as usability, features, cost, etc. Interestingly, as we found in our Android case study, apps with similar functionalities can have a different performance. For instance, to visit an article in Wikipedia, the browser Chrome uses more power and transmits more data over the network than the browser Opera (mini). However, the former uses less CPU. It means that there exist a trade-off in terms of performance between different apps. The same keeps for any category of apps in any marketplace and mobile device platform. But performance metrics alone are not important, as performance depends on the context of use and what is important to the end-user~(\cite{falaki_diversity_2010}). \comment{For example, a majority of users in underdeveloped markets face constraints not commonly seen in developed markets: high costs when data connections are available, low-end devices with reduced memory, and
limited opportunities to recharge batteries during the day\footnote{https://developer.android.com/distribute/best-practices/develop/build-for-the-next-billion.html
}.
To address the needs of users, app performance metrics must be aligned closely with users preferences.} Thus, if the user is using a Wi-Fi network connection, battery life could be more important than network usage, but while driving during a trip abroad the preference is different because of cost of the data is higher and mobile car chargers are cheap.

Making performance information available regarding power, CPU, memory, and network usages would be useful for both mobile users and developers. 
It would put pressure on developers to build more efficient apps, which benefits final users. However, even if information about performance metrics of mobile device apps would be available in marketplaces, the selection of optimal apps would be complicated because of the cognitive effort imposed to discriminate between different sets of apps and different metrics. This is what we define as the \textsl{App Selection Problem} (ASP). 


In a recent research we presented a simple recommendation system to propose optimal sets of Android apps that minimize power and data usages while maximizing the apps rating (\cite{saborido_optimizing_2016}). 
In this paper we propose APOA (an App Performance Optimization Advisor) that could complement existing mobile device marketplaces allowing users and developers to compare apps in terms of any combination of metrics (power, CPU, memory, and network usages, and rating). 
Each of the possible combinations of these metrics is a particular instance of the ASP that matches a context of use (travel abroad, driving, ...). Therefore, the resolution of an instance of the ASP is a recommendation of optimal apps for a given context. 
The main contribution of this paper is threefold:
\begin{itemize}

\item{The ASP is introduced and formally defined.}

\item{APOA is proposed as a recommendation system for mobile device marketplaces and it is evaluated over a subset of 140 Android apps and different contexts of use.}

\item{APOA's implementation is publicly available for download\footnote{http://www.ptidej.net/downloads/replications/APOA/}.}

\end{itemize}

The rest of the paper is organized as follows. \mbox{Section \ref{sec:selectionproblem}} formally defines the ASP based on different metrics and introduces basic concepts related to optimization. \mbox{Section \ref{sec:contexts}} defines four different contexts of use which are used to evaluate our approach. \mbox{Section \ref{sec:approach}} explains and describes APOA, our App Performance Optimization Advisor. \mbox{Section \ref{sec:casestudy}} shows the case study based on a subset of Android apps used to evaluate APOA. Next, \mbox{Section \ref{sec:results}} shows the evaluation of APOA and the benefits of using it. \comment{W}e discuss about threats to validity of our study and performance measures in \mbox{Section \ref{sec:threats}} and \mbox{Section \ref{sec:discussion}}, respectively. \comment{Section \ref{sec:relatedWork} summarizes related work} and we conclude the paper in \mbox{Section \ref{sec:conclusion}}.


\section{The App Selection Problem (ASP)}
The ASP is when a user wants to select an app or a set of apps to achieve a goal or task but is faced with numerous ratings, reviews, performance information, and context. For instance the right bandwidth-saving app to help you look up restaurant reviews when you are paying \$1 dollar per kilobyte might not be optimal later when you are on bandwidth abundant Wi-Fi in your hotel room.

Given the huge number of available apps in mobile device apps marketplaces, the number of existing categories, and taking into account that, in a category, apps often share similar functionalities, we define the selection of apps as a combinatorial problem. Let $\mathcal{C}=\{C_1, ..., C_N\}$ be a set  of $N$ categories. Further, assume that each
category $C_i$ contains a set $A_i$ of apps.  
An  element
$\textbf{x}$ of the search space  $\textit{F}$, \mbox{$\textbf{x}= (x_1, \dots, x_N)$}, is
a set of apps where $x_l$ is an app selected from $A_l$ (with $l \in \left\lbrace 1, \dots, N \right\rbrace$). A solution $\textbf{x}$ contains one app from each category in $\mathcal{C}$. Considering the previous, the size of the search space is given by $\prod_{i=1}^{N}|A_{i}|=|A_{1}|\cdot |A_{2}|\cdots |A_{N}|$, where the operator $|B|$ represents the number of elements in a set $B$. 

We consider that an app has an associated cost in terms of performance metrics and, therefore, a combination of apps also has it. However, not all the possible combinations of apps are valid, because some of them could be more efficient than others. In addition to performance metrics, we consider the rating of apps as a ``quality'' indicator of mobile device apps because, 
as it was studied by \cite{harman_app_2012}, apps ratings are highly correlated with the number of app downloads, which is a measure of their success. For this reason we use five metrics (four performance metrics to minimize and a quality metric to maximize) which are enumerated next:

\begin{itemize}
\item {Energy consumption and power usage. 
Energy consumption determines the battery life of mobile devices and, therefore, their availability. Without energy a mobile device cannot be operated. Energy is defined as the capacity of doing work while power is the rate of using energy. Energy ($E$) is measured in Joules ($J$) while power ($P$) is measured in Watts ($W$). Energy is equal to power times the time period $T$ in seconds. Therefore, $E = P \times T$. Thus, if an app uses two Watts of power for five seconds it consumes 10 Joules of energy. Since power usage is hard to be interpreted for users, we translate it into battery life, which specifies the duration of the battery in hours. We describe battery life as it is shown in \mbox{Equation \ref{eq:batterylife}}, where $Load$ is the average power usage of a load (for us an app or set of apps), and $C_B$ and $V_B$ are the electric charge (in $Ah$) and voltage (in $V$), respectively, of the phone's battery.

\begin{equation}
\label{eq:batterylife}
Battery_{Life} = \frac{C_B \times V_B}{Load}
\end{equation}}

\item {CPU usage describes the proportion of time that the processor is in use. A mobile device's CPU usage can vary depending on the types of tasks that are being performed by an app. We measure CPU usage in percentage (\%), which indicates how much of the processor's capacity is currently in use by the system. Typically CPU is one of the primary sources of energy consumptions \comment{(\mbox{\cite{halpern_mobile_2016}}).}}

\item {Memory usage. It is the amount of memory (RAM) an app uses when it is running. This memory is used to save  internal data and app's instructions. We measure memory usage in megabytes ($MB$). Memory limits the number of apps we can run and the amount of data they can work with, for this reason this metric is considered. 
}

\item {Network usage. It refers to the amount of data moving across a network (Wi-Fi, 3G, 4G, $\dots$). We measure network usage in megabytes ($MB$). 
We consider this metric because network access could be expensive in terms of bandwidth costs.}

\item{Rating. It is a number between one and five associated to mobile device apps. \comment{The rating of an app in a marketplace is calculated from its user ratings.}}
\end{itemize}

The rating is available in marketplaces but performance metrics are not and thus it has to be calculated or estimated.
Once these metrics are collected and available, the ASP is modeled as an optimization problem using as objective function any combination of the described metrics. A $k$-combination of a set $S$ is a subset of $k$ distinct elements of $S$. If the set has $n$ elements, the number of $k$-combinations is equal to the binomial coefficient $\binom{n}{k} = \frac{n!}{k! \cdot (n-k)!}$ whenever $k \leq n$ and which is zero when $k > n$. Given that we consider five metrics ($n=5$) and we are interested in any combination of them ($k \in \left\lbrace 1,2,3,4,5 \right\rbrace$), there exist $\binom{5}{1} + \binom{5}{2} + \binom{5}{3} + \binom{5}{4} + \binom{5}{5} = 31$ different combinations, which can be considered as instances of the ASP. We summarize all of them in \mbox{Table \ref{tab:problems}}. The first column contains an identifier associated with each instance. Performance metrics and the rating are shown from the second to the sixth column, respectively. In these columns the symbol ``$\circ$'' means that the corresponding metric is considered in a particular instance of the problem. On the contrary, the ``-'' symbol means that the corresponding metric is not considered. Finally, the last column specifies the number of objective functions of each instance. Thus, \mbox{Instance 1} considers power usage as metric to be optimized while \mbox{Instance 31} considers power, CPU, memory, and network usages, and the rating.

\begin{table}[!htbp]
\centering
\caption{Instances of the ASP when up to five different metrics are considered. 
}
\label{tab:problems}
\begin{scriptsize}
\begin{tabular}{ccccccc}
\hline
Instance & Power & CPU & Memory & Network & Rating & \#Obj \\
\hline
1       & $\circ$     & -   & -      & -       & -      & 1          \\
2       & -     & $\circ$   & -      & -       & -      & 1          \\
3       & -     & -   & $\circ$      & -       & -      & 1          \\
4       & -     & -   & -      & $\circ$       & -      & 1          \\
5       & -     & -   & -      & -       & $\circ$      & 1          \\
6       & $\circ$     & $\circ$   & -      & -       & -      & 2          \\
7       & $\circ$     & -   & $\circ$      & -       & -      & 2          \\
8       & $\circ$     & -   & -      & $\circ$       & -      & 2          \\
9       & $\circ$     & -   & -      & -       & $\circ$      & 2          \\
10      & -     & $\circ$   & $\circ$      & -       & -      & 2          \\
11      & -     & $\circ$   & -      & $\circ$       & -      & 2          \\
12      & -     & $\circ$   & -      & -       & $\circ$      & 2          \\
13      & -     & -   & $\circ$      & $\circ$       & -      & 2          \\
14      & -     & -   & $\circ$      & -       & $\circ$      & 2          \\
15      & -     & -   & -      & $\circ$       & $\circ$      & 2          \\
16      & $\circ$     & $\circ$   & $\circ$      & -       & -      & 3          \\
17      & $\circ$     & $\circ$   & -      & $\circ$       & -      & 3          \\
18      & $\circ$     & $\circ$   & -      & -       & $\circ$      & 3          \\
19      & $\circ$     & -   & $\circ$      & $\circ$       & -      & 3          \\
20      & $\circ$     & -   & $\circ$      & -       & $\circ$      & 3          \\
21      & $\circ$     & -   & -      & $\circ$       & $\circ$      & 3          \\
22      & -     & $\circ$   & $\circ$      & $\circ$       & -      & 3          \\
23      & -     & $\circ$   & $\circ$      & -       & $\circ$      & 3          \\
24      & -     & $\circ$   & -      & $\circ$       & $\circ$      & 3          \\
25      & -     & -   & $\circ$      & $\circ$       & $\circ$      & 3          \\
26      & $\circ$     & $\circ$   & $\circ$      & $\circ$       & -      & 4          \\
27      & $\circ$     & $\circ$   & $\circ$      & -       & $\circ$      & 4          \\
28      & $\circ$     & $\circ$   & -      & $\circ$       & $\circ$      & 4          \\
29      & $\circ$     & -   & $\circ$      & $\circ$       & $\circ$      & 4          \\
30      & -     & $\circ$   & $\circ$      & $\circ$       & $\circ$      & 4          \\
31      & $\circ$     & $\circ$   & $\circ$      & $\circ$       & $\circ$      & 5        \\
\hline
\end{tabular}
\end{scriptsize}
\end{table}

Considering that different apps belong to $N$ different categories in the marketplace and that a solution $\textbf{x}$ of the ASP is a combination of apps, metrics to be optimized can be calculated as follow:

\begin{equation}
\label{eq:power}
power(\textbf{x}) = \frac{\sum_{i=1}^{N}{power(x_i)}}{N}
\end{equation}

\begin{equation}
\label{eq:cpu}
CPU(\textbf{x}) = \frac{\sum_{i=1}^{N}{CPU(x_i)}}{N}
\end{equation}

\begin{equation}
\label{eq:memory}
memory(\textbf{x}) = \frac{\sum_{i=1}^{N}{memory(x_i)}}{N}
\end{equation}

\begin{equation}
\label{eq:network}
network(\textbf{x}) = \frac{\sum_{i=1}^{N}{network(x_i)}}{N}
\end{equation}

\begin{equation}
\label{eq:rating}
rating(\textbf{x}) = \frac{\sum_{i=1}^{N}{rating(x_i)}}{N}
\end{equation}

\noindent  In Equations (\ref{eq:power}), (\ref{eq:cpu}), (\ref{eq:memory}), and  (\ref{eq:network}),   $power(x_i)$, $CPU(x_i)$, $memory(x_i)$,  and
$network(x_i)$ are the average values of power (in W), CPU usage (in \%), memory usage (in MB), and network usage (in MB)
for app $x_i$ in a certain number of runs and for a given number of exercised app functionalities.
In  Equation (\ref{eq:rating}),
$rating(x_i)$ is  the rating  of the application  $x_i$ in the marketplace. Notice that the constant $N$ is just a rescaling factor and thus, in this case, optimizing $\frac{\sum_{i=1}^{N}{rating(x_i)}}{N}$ is the same as optimizing $\sum_{i=1}^{N}{rating(x_i)}$.
The same holds for performance metrics.

Solving any instance of the ASP finds a set of optimal apps which maximize the quality metric (rating) and--or minimize performance metrics\footnote{Minimizing power usage is equivalent to maximize battery life.}. If only one metric is optimized, the problem is considered as a single objective optimization problem. On the contrary, if the number of metrics to optimize is greater than one, the problem is considered as a multi-objective optimization problem. We think that users likely prefer to optimize more than one metric, for example maximizing the rating while at least one performance metric is also optimized. Consequently we are more interested in instances of the ASP in which two or more metrics are involved. Therefore, we focus on multi-objective optimization problems.

\section*{Background in Multi-objective Optimization}
\label{sec:background}
%
%
Many real-world problems involve dealing with several conflicting criteria or objectives, which must be optimized
simultaneously. These problems, called multi-objective optimization problems, are defined by criteria and
constraints that are usually expressed through mathematical functions. Since, in general, it is impossible to find a solution where all the objectives can reach their individual
optima at the same time, it is necessary to identify compromise solutions, which are so-called Pareto optimal
or efficient solutions, where none of the objectives can achieve a better value without getting worse in at least one of the other objective function values. The set of Pareto optimal solutions is called the Pareto
optimal set and its image in the objective space is known as the Pareto optimal front.

Formally multi-objective optimization problems are mathematical programming problems with a vector-valued objective function, which is usually denoted by
${\textbf{f}(\textbf{x})=(f_1(\textbf{x}),...,f_m(\textbf{x}))}$, where $f_j(\textbf{x})$, for $j=1,\ldots,m$, is a real-valued function defined on the feasible region
$\textit{F} \subseteq \Re^M$. Consequently, the decision space belongs to $\Re^M$ while the criterion space belongs to $\Re^m$, and the multi-objective optimization problem can be stated as follows:
\begin{displaymath}
\begin{array} {ll}
\mathrm{optimize} & [f_1(\textbf{x}),...,f_m(\textbf{x})]   \\
\mathrm{s.t.} & \textbf{x} \in \textit{F} \\
\end{array}
\label{eq:mop}
\end{displaymath}

In the functional space of criterion, some objective functions should be maximized ($j \in J_{max}$) while others should be minimized ($j \in J_{min}$), these subsets of indices verifying that $J_{max} \cup J_{min}=\{1,...,m\}$. In this context, optimality is defined on the basis of the concept of \textit{dominance}, in such a way that solving the above problem implies finding the subset of \textit{non-dominated solutions}, that is those feasible solutions which are not dominated by any other feasible one. A feasible solution $\textbf{x}^0$ \textit{dominates} another solution $\textbf{x} \in \textit{F}$ if and only if $f_j(\textbf{x}^0) \geq f_j(\textbf{x})$, for every $j\in J_{max}$ and $f_j(\textbf{x}^0)\leq f_j(\textbf{x})$, for every $j\in J_{min}$, with at least one strict inequality. The set of non-dominated solutions will also be referred by \textit{Pareto optimal solutions} and define the efficient frontier or \textit{Pareto optimal front} of the multi-objective optimization problem (see, for instance, \cite{Miettinen1999}). 

Different methodologies exist to solve multi-objective optimization problems. Multiple Criteria Decision Making (MCDM) (\cite{HwangMasud,Miettinen1999}) and Evolutionary Multi-objective Optimization (EMO) (\cite{Deb_EMObook,Coello_EMObook}) are the most popular methodologies which have contributed with several different approaches to solve real problems. EMO algorithms, in general, find a evenly distributed set of Pareto optimal solutions to approximate the Pareto optimal front while MCDM takes into account some user preferences to find a reduced set of optimal solutions. 
 


\label{sec:selectionproblem}

\section{Contexts of Use}
\label{sec:contexts}
Mobile device users could need different kinds of performance depending on their location, time, and intent.
There are many different contexts of mobile device use due to the wide variation among users' usage (\cite{falaki_diversity_2010}).
We claim that, depending on the context, some performance metrics are more important than others. Therefore, even if different apps offer similar functionalities, one app could be preferred over others because of its performance in that context. Thus, we consider that the context of use affects users' preferences about the metrics to be optimized. 
In this section, we define four different contexts of use which are associated to instances of the ASP. Next, we present these contexts of use which are  used later to evaluate APOA.

\subsection*{Travel Abroad.}

This context considers users who travel abroad, whether for work or leisure. In that case, we suppose that \comment{the} most important performance metrics are battery life \comment{(}as the time between consecutive charges is likely longer\comment{)} and network usage \comment{(}as  data roaming is usually expensive\comment{)}. This context of use corresponds to \mbox{Instance 8} of the ASP.

\subsection*{Old Devices.}

In addition to energy consumption and network usage, CPU and memory usages are also important metrics because apps that are CPU and--or memory greedy, slow down devices, impacting negatively the \comment{user experience}. This is \comment{e}specially important for mobile users \comment{in emerging markets} who \comment{usually} own old mobile devices. 
In this context, we consider that the most important metrics are CPU and memory usages. Therefore, this context of use corresponds to \mbox{Instance 10} of the ASP.

\subsection*{Driving.}

This context includes users who use their mobile devices as GPS navigator. At this point we divide this context in two different sub-contexts.

\begin{enumerate}
\item{Phone is not plugged into the car charger. Here energy consumption is the most important performance metric to be optimized. This context of use corresponds to \mbox{Instance 1} of the ASP.}

\item{Phone is plugged into the car charger. In this case, we consider that the most important performance metrics are the network usage (to minimize data plan usage), and CPU and memory usages (to avoid lags in GPS navigation due to the phone's slowing down). This context of use corresponds to \mbox{Instance 22} of the ASP.}
\end{enumerate}

These contexts of use and their correspondences to different instances of the ASP are summarized in \mbox{Table \ref{tab:contexts}}. The first column contains contexts of use while the second one indicates the associated instance of the ASP. Performance metrics are shown from the third to the sixth column. In these columns the symbol ``$\circ$'' means that the corresponding performance metric is considered in that particular instance of the problem, as it was shown in \mbox{Table \ref{tab:problems}}. On the contrary, the ``-'' symbol means that the corresponding performance metric is not considered.

\begin{table}[!htbp]
\centering
\caption{Correspondence between contexts of use and instances of the ASP.}
\label{tab:contexts}
\begin{scriptsize}
\begin{tabular}{lccccc}
\hline
Context of Use & Instance & Power & CPU & Memory & Network      \\
\hline
Driving (1)  & 1        & $\circ$     & -   & -      & - \\
Driving (2)    & 22       & -     & $\circ$   & $\circ$      & $\circ$ 
\\
Old devices & 10       & -     & $\circ$   & $\circ$      & -           \\
Travel abroad & 8        & $\circ$     & -   & -      & $\circ$           \\
\hline
\end{tabular}
\end{scriptsize}
\end{table}

\section{APOA: An App Performance Optimization Advisor}

\begin{figure*}[!htbp]
\centering
\includegraphics[scale=0.55]{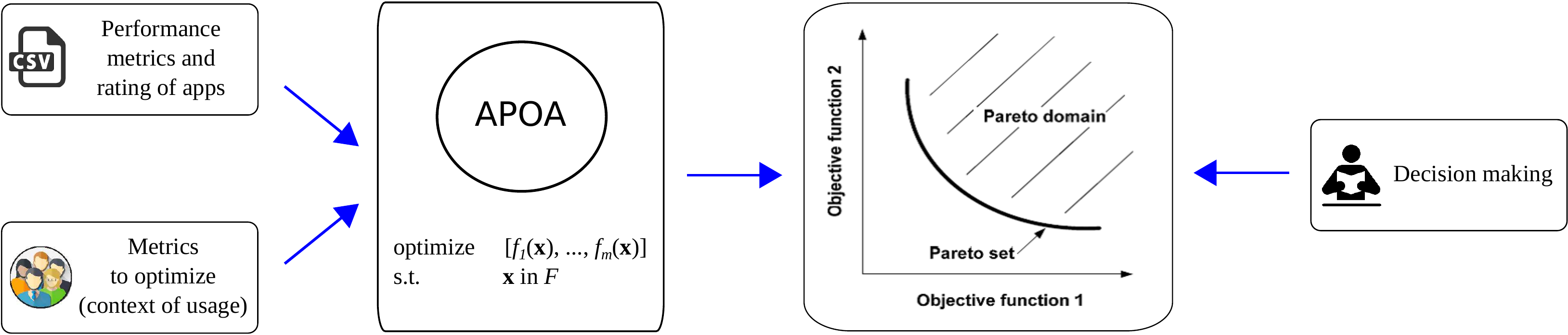}
\caption{APOA conceptual sequence of steps. It uses as input a set of metrics to optimize and metric values for a set of apps belonging to different categories. It solves an optimization problem and it generates, as output, a Pareto optimal front. Each solution in the Pareto optimal front represents an optimal set of apps. Over the resulting Pareto optimal front the user selects the most preferred solution.}
\label{fig:approach}
\end{figure*}

The  APOA  process is shown in \mbox{Figure
\ref{fig:approach}}. APOA uses as input the set of metrics to be
optimized, which defines the context of use, and metric values of mobile device apps (this data can be given as a comma separated values (CSV) file). Using this information
it solves the corresponding instance of the ASP generating, as output, a Pareto optimal set of apps over which the user chooses the most preferred solution (decision making). If the input contains metrics for apps in a single specific category APOA will found optimal apps in that category. On the contrary, if metrics are given for sets of apps in different categories, our approach will found optimal combinations of apps. APOA could be considered as a solver of the ASP \comment{and it is transparent to the data collection process}.

Given a set of apps in a category, not all of them have similar metrics. For instance, if a user is interested in installing a popular and energy efficient camera app, not all the existing camera apps in the marketplace should be considered because some of them could be less popular and more energy greedy than others in the same category. In that case the number of apps to take into account could be reduced, because only energy efficient and popular apps should be taken into account. It means that only camera apps which are Pareto equivalent considering these two metrics, rating and battery life, should be shown to the user. If we extend this fact to all the categories the search space can be reduced removing Pareto dominated apps in each category. APOA always applies this reduction approach, which is formally described in \mbox{Algorithm \ref{alg:spaceReduction}}. As it was commented in \mbox{Section \ref{sec:selectionproblem}}, $A_i$ is the set of apps contained in category $C_i$, with $i \in \left\lbrace 1,\dots,N \right\rbrace$.

\begin{algorithm} [!htbp] \caption{Search space reduction.}
\begin{footnotesize}
\begin{algorithmic}[1]
\Require Metrics to optimize and apps' metrics.
\Ensure Pareto optimal apps in each category.
\ForEach {$A_i$ where $i \in \left\lbrace 1,\dots,N \right\rbrace$}
\State $A'_i=$ Apply Pareto dominance to apps in $A_i$.
\EndFor
\\
\Return $A'_i \ \forall i \in \left\lbrace 1,\dots,N \right\rbrace$
\end{algorithmic}
\end{footnotesize}
\label{alg:spaceReduction}
\end{algorithm}

After the search space reduction, depending on the number of categories and apps by category, the decision space could still be huge. In that case, 
EMO algorithms as \mbox{NSGA-II}~(\cite{DebNSGAII}) can be used to generate a set of close to optimal solutions. On the contrary, if the search space is small, an exhaustive search can be applied. In this case APOA enumerates all the possible combinations and applies the Pareto dominance relation to select the Pareto optimal combinations of apps.

\mbox{Algorithm \ref{alg:exhaustiveSearch}} shows the exhaustive search used by APOA.
First, it generates (line 1) all the possible combinations of apps belonging to each category. It means that, if there are $N$ categories and each category contains a set of $A'_i$ apps after the search space reduction, \mbox{$\prod_{i=1}^{N}|A'_{i}|$}  different combinations of apps are generated.  Second, for each combination, objective values associated to the metrics to be optimized are calculated using equations (\ref{eq:power}), (\ref{eq:cpu}), (\ref{eq:memory}), (\ref{eq:network}), and--or (\ref{eq:rating}) (line 2). Third, the Pareto dominance relation is applied over the combinations of apps to select the Pareto optimal ones (line 3).  

\begin{algorithm} [!htbp] \caption{Exhaustive search.}
\begin{footnotesize}
\begin{algorithmic}[1]
\Require Metrics to optimize and apps' metrics.
\Ensure Pareto optimal set of apps (the Pareto optimal front).
\State $Comb$ = Combinations of apps belonging to each category.
\State Calculate objective functions of each combination in $Comb$.
\State $output=$ Apply Pareto dominance to $Comb$.
\\
\Return output
\end{algorithmic}
\end{footnotesize}
\label{alg:exhaustiveSearch}
\end{algorithm}

As output, APOA shows a Pareto optimal front and the existing trade-off for each solution and optimized metrics. All the solutions in a Pareto optimal front are equivalents considering the Pareto dominance relation, because if an objective is improved another objective is worsened. For this reason there is not a best solution and the trade-off between different solutions and objectives should be analyzed by the user or decision maker (DM) to choose the most preferred solution (decision making). The trade-off of a solution and an objective function specifies the difference in percentage with respect to the best value of this objective function in the Pareto optimal front. APOA returns this information in a table enumerating all the Pareto optimal solutions and their trade-offs. In addition, to make easier the comparison between different solutions APOA shows trade-offs in a stacked bar graph. The bars in a stacked bar are divided into different categories, one for each optimized metric. 

APOA and instances of the ASP has been implemented in jMetal, an object-oriented Java-based framework for multi-objective optimization (\cite{Durillo2011}). An additional input parameter is used to specify if APOA solves an instance of the ASP performing an exhaustive search or running the \mbox{NSGA-II} algorithm.
\label{sec:approach}

\section{Case Study}
\label{sec:casestudy}
In order to evaluate APOA we propose a case study over a subset of Android apps. We choose the Android ecosystem because it is the most popular mobile operative system globally. The goal of this study is to assess APOA capabilities with the purpose of understanding
APOA applicability to find optimal sets of apps in terms of different metrics. 
We want to balance app qualities to find the best set of apps given a
context. This subset of apps will save users effort and time in evaluating apps.
The evaluation is executed from the perspective of users who wish to select and
install a set of apps from a set of categories, and the perspective \comment{of} developers who need to benchmark their apps against apps in the same category. 

The context of the case study consists of a subset of Android apps
belonging to different categories in the Google Play
marketplace. First, we select the most popular Android apps in the marketplace (\mbox{Subsection \ref{subsec:selectionofapps}}). Second, we define typical usage scenarios (\mbox{Subsection \ref{subsec:scenarios}}). Finally, we collect and process data of different metrics for the selected apps (\mbox{Subsection \ref{subsec:datacollection}}). 
In the following subsections we detail how we handle each of these steps. Using all of these measures, APOA is evaluated considering different contexts of use. 
In our experiments, we use a LG Nexus 4 Android phone, equipped with Android Lollipop operating system (version 5.1.1, Build number LMY47V). Battery values of this phone's battery are $C_B=2.10Ah$ and $V_B=3.8V$, which are used to translate power usage to battery life.


\subsection{Selecting Most Popular Android Apps}
\label{subsec:selectionofapps}
We select a set of free Android apps belonging to seven different categories ($N = 7$). These apps are chosen considering the number of downloads in Google Play, because 
Android users choose apps that others choose. Based on the categories
used by \cite{saborido_optimizing_2016}, we define a subset of these categories considering the most common apps used by mobile device users: browsers, cameras, flash lights, music players, news viewers, video players, and weather forecast. For each category \comment{the 100 applications with the best rating} were selected and their descriptions, statistics, and \texttt{apk} files were downloaded automatically using a Perl script and the tool Play Store Crawler\footnote{https://github.com/Akdeniz/google-play-crawler}.
 In addition, a stress-test Python script which uses the \texttt{adb}\footnote{http://developer.android.com/tools/help/adb.html} 
and Monkey\footnote{http://developer.android.com/tools/help/monkey.html} 
Android tools was used to remove the apps that crashed during their execution on the real phone. This stress-test was similar to the approach proposed by \cite{Li14-icsme}, configuring Monkey to generate 180 random events during 60 seconds (three events per second). Around $2\%$ of the apps crashed during this test and, therefore, were removed. Considering the rest of apps, the subset of most downloaded 20 applications were finally selected for each category. Therefore, in total, we analyze 140 apps for all of the seven selected categories. 

\subsection{Definition of Typical Usage Scenarios}
\label{subsec:scenarios}
For each app in a category, we propose a typical usage scenario and we play it automatically while performance metrics are measured.
We use the scenarios defined by \cite{saborido_optimizing_2016}, which were collected interacting with each app under study using the Android app HiroMacro\footnote{https://play.google.com/store/apps/details?id=com.prohiro.macro}. 
This software allows to generate scripts containing the \texttt{touch} and \texttt{move} events while a user interacts with each app directly on the phone. The resulting script can be played automatically using the same app but it was converted to a Monkeyrunner format. Thus, the interaction to collect the scenario is done using the phone and the actions can be played automatically from our code using the Monkeyrunner Android tool. 
For all of the apps, a usual scenario for users (for example, navigating using the browser, taking some pictures, or playing a song) was simulated. The scenarios defined for each category and collected for each app are summarized in \mbox{Table \ref{table:scenarios}}.


\begin{table}[H]
\centering
\caption{Summary of typical usage scenarios defined for each app category.}
\label{table:scenarios}
\begin{scriptsize}
\begin{tabular}{ll}
\hline
Category & Scenario description                                                  \\
\hline
Browsers    & Search and read an article in Wikipedia.\\
Cameras     & Take three pictures. \\
Flash Lights & Use the torch during 10 seconds. \\
Music Players & Play two songs during 20 seconds. \\
News & Read the two first news. \\
Video Players & Play a movie for 30 seconds. \\
Weather  & Get the forecast for two different cities.
\\
\hline
\end{tabular}
\end{scriptsize}
\end{table}

\subsection{Data Collection and Processing}
\label{subsec:datacollection}
For each app we run a simple scenario to start the app, skip the initial tutorial (if it exists), and interact with the app to simulate the user interaction. These scenarios are run automatically while performance metrics are collected. We run each app 20 times and, in each run, the app is uninstalled after its usage and the cache is cleaned using the Android command \texttt{adb}. A description of the steps is given in Algorithm~\ref{alg:script} which has been implemented as a Python script. For simplicity we include all the performance metrics in the same script but, in fact, power usage is collected individually to avoid any impact of other metrics' measurements on it. As it is described, all apps are executed before a new run is started to avoid that the cache memory on the phone stores information related to the app.

\begin{algorithm} [!htbp] \caption{Collecting performance metrics.}
\begin{footnotesize}
\begin{algorithmic}[1]
\Require categories, number of runs, and list of apps
\ForEach {category}
\ForEach {run}
\ForEach {app}
\State Install app (using \texttt{adb}). 
\State Run app (using \texttt{adb}). 
\State Wait 10 seconds (to load the app fully). 
\If {(app requires initialization)} 
\State Play set-up (using Monkeyrunner). 
\EndIf
\State Start \texttt{top} command (using \texttt{adb}). 
\State Start \texttt{dumpsys} command (using \texttt{adb}). 
\State Start \texttt{tcpdump} (using \texttt{adb}). 
\State Start oscilloscope to measure power usage. 
\State Play scenario to simulate user interaction. 
\State Stop oscilloscope. 
\State Stop \texttt{tcpdump} (using \texttt{adb}). 
\State Stop \texttt{dumpsys} command (using \texttt{adb}). 
\State Stop \texttt{top} command (using \texttt{adb}). 
\State Stop app (using \texttt{adb}).
\State Uninstall app (using \texttt{adb}). 
\State Download the \texttt{tcpdump} file (using \texttt{adb}). 
\State Download file containing memory usage (using \texttt{adb}). 
\State Download file containing CPU usage (using \texttt{adb}). 
\EndFor
\EndFor
\EndFor
\end{algorithmic}
\end{footnotesize}
\label{alg:script}
\end{algorithm}


Before the experiments, the screen brightness is set to the minimum value and the phone is set to keep the screen on. In order to avoid any kind of interference during the measurements, only the essential Android services are run on the phone. Because some apps need to be set up before they can be used, in Algorithm \ref{alg:script} there is a step to initialize the app when it is required.
In these cases, the initialization process uses Monkeyrunner to run a sequence of events to set up the app. These events were collected previously for each app using HiroMacro as was explained before in this section. Finally, when the data are collected, they are processed to calculate and save in a CSV file the associated rating existing in Google Play and the average power, CPU, memory, and network usages for each app over all the runs. 

Power usage is measured using the same approach proposed by \cite{saborido_optimizing_2016} that uses a digital oscilloscope TiePie Handyscope HS5\footnote{http://www.tiepie.com/en/products/Oscilloscopes/Handyscope\_HS5} 
which offers the LibTiePie SDK. 
The mobile phone is powered by a power supply and, between both we connect, in series, a uCurrent\footnote{http://www.eevblog.com/projects/ucurrent/} device, 
which is a precision current adapter for multimeters converting the input current in a proportional output voltage ($V_{out}$). Knowing the input current ($I$) and the voltage supplied by the power supply ($V_{sup}$), we use the Ohm's Law to calculate the power usage ($P$) as $P = V_{sup} \cdot I$. \comment{Although from Android API level 21 Google made available new APIs to estimate the energy consumption of apps, this hybrid software/hardware based approximation, using statistics from the battery itself, is still 5\% inaccurate (\mbox{\cite{di_nucci_software-based_2017}}). Instead we used the described hardware based approach to get more precise energy measurements.} 

CPU usage is collected using the same approach used by \cite{gui_truth_2015}. Specifically, the \texttt{top} command is run on the phone in background obtaining the percentage of CPU usage associated to an Android app. Every second, this information is added to a file stored on the phone.

Memory usage is measured using the \texttt{dumpsys meminfo} command on the phone. Every second, this information is obtained for the process associated to the Android app and it is added to a file stored on the phone. Specifically we measure memory using the Proportional Set Size (PSS), which is proposed in the Android documentation.
This is a measurement of the app's RAM use that takes into account sharing pages across processes. 
This metric is different from the one used by \cite{gui_truth_2015}, where authors use the Resident Set Size (RSS). RSS indicates how many physical pages are associated with the process, which is less precise. 

Network usage is collected using the tool \texttt{tcpdump}\footnote{http://www.androidtcpdump.com/}, 
which has been used in \comment{previous} works (\cite{gui_truth_2015,saborido_optimizing_2016}). \texttt{tcpdump} is a command line packet capture utility, useful for capturing packets from the Wi-Fi and cellular connections. We use this tool via \texttt{adb} to capture the number of bytes transmitted over the network connection while an app is running. 

In total, for this case study, we collected 11,200 files (more than one terabyte of data).

\section{Results}
\label{sec:results}
The rating, existing in the marketplace, and the collected performance metrics of the 140 Android apps selected for the case study are used in this section to evaluate APOA. First, we run APOA over the collected data to solve all the instances of the ASP. Second, we check using APOA if popular apps are optimal in terms of performance metrics. Third, we use APOA to recommend optimal sets of Android apps for contexts of use described in \mbox{Section \ref{sec:contexts}}. Finally, we show how collected data about performance metrics can be used to assist developers. 

\subsection{Resolution of the App Selection Problem}
In this subsection we present the results of using APOA to solve all the instances of the ASP for the Android case study. We use both the exhaustive search and the EMO algorithm \mbox{NSGA-II}. Experiments are run in a Lenovo ThinkPad laptop ($4 \times $ Intel Core i5-6200U CPU @ $2.30GHz$) running Debian GNU/Linux Stretch. \mbox{Table \mbox{\ref{table:NSGAIIparameters}}} shows the parameters and operators used in \mbox{NSGA-II} which are commented by \cite{Deb_EMObook}. The Single Point crossover operator is used because it is one of the simplest crossover operators and it works reasonably well in combinatorial problems. When two parents are selected, with a probability of $P_x$ the operator is used to create new individuals. It selects a point on both parents and all data beyond that point in either individual is swapped between the two parents. The resulting solutions or individuals are the offspring. Considering the mutation, the flip mutation operator is used. It changes the value of a gene in the individual, with a probability of $P_m$, with a new value generated randomly in the lower and upper bounds range. The binary tournament selection operator is used to select individuals in the population to create the offspring. This operator selects two solutions randomly in the population and chooses the best one, or one of them with a probability of $0.5$ if they are equivalents.

\begin{table}[!htbp]
\centering
\caption{Parameters settings for the EMO algorithm \mbox{NSGA-II}.}
\label{table:NSGAIIparameters}
\begin{scriptsize}
\begin{tabular}{ll}
\hline
Parameter             & Value                   \\
\hline
Population size       & 200                     \\
Generations           & 300                     \\
Crossover operator   & Single point crossover \\
Crossover probability ($P_x$) & 0.9 \\
Mutation operator     & Flip mutation         \\
Mutation probability ($P_m$)  & $1/C=0.125$ \\
Selection operator    & Binary tournament
\\
\hline    
\end{tabular}
\end{scriptsize}
\end{table}

\mbox{Table \ref{tab:solutions}} shows the number of solutions obtained by APOA running the exhaustive search. Each row corresponds to each instance of the ASP. From the third to the ninth columns the number of Pareto optimal apps in each category is shown. The tenth column shows the number of possible combinations of apps (solutions) considering all the categories. Finally, the last column, specifies the number of Pareto optimal solutions over all the possible solutions (previous column). As it is shown, after the search space reduction the number of optimal apps in each category is reduced. For example, if we consider Instance~31 and the browsers category, out of 20 apps nine (45.00\%) are Pareto optimal. It means that 11 apps are discarded and they are not considered in the optimization process. For this reason, after the search space reduction the initial number of possible solutions is reduced to the value indicated in the tenth column. Over all of these solutions a reduced subset are Pareto optimal (eleventh column). If we consider again \mbox{Instance 31}, 23,591 (0.15\%) over 15,459,444 existing solutions are Pareto optimal. This shows the need of a recommendation system as APOA to filter out non-optimal solutions and help users to reduce the cognitive effort to choose the most preferred one.

\begin{table*}[!htbp]
\centering
\caption{Number of solutions and Pareto optimal solutions for each instance of the ASP for the Android case study.}
\label{tab:solutions}
\begin{scriptsize}
\begin{tabular}{ccccccccccc}
\hline
Instance & \#Obj & Browsers & Cameras & Flash Lights & Music Players & News & Video Players & Weather & Solutions & Optimal \\
\hline
1 & 1       & 1        & 1       & 1           & 1            & 1    & 1            & 1       & 1         & 1       \\
2 & 1       & 1        & 1       & 1           & 1            & 1    & 1            & 1       & 1         & 1       \\
3    & 1    & 1        & 1       & 1           & 1            & 1    & 1            & 1       & 1         & 1       \\
4       & 1 & 1        & 1       & 1           & 1            & 1    & 1            & 1       & 1         & 1       \\
5      & 1  & 2        & 1       & 1           & 1            & 1    & 1            & 2       & 4         & 4       \\
6 & 2       & 4        & 3       & 2           & 5            & 3    & 3            & 2       & 2,160      & 48      \\
7 & 2       & 3        & 1       & 1           & 3            & 2    & 4            & 4       & 288       & 18      \\
8 & 2       & 1        & 5       & 1           & 3            & 1    & 2            & 4       & 120       & 27      \\
9  & 2      & 2        & 2       & 3           & 4            & 2    & 3            & 2       & 576       & 15      \\
10 & 2      & 2        & 4       & 2           & 4            & 1    & 1            & 5       & 320       & 31      \\
11 & 2      & 5        & 2       & 2           & 5            & 3    & 4            & 2       & 2,400      & 70      \\
12 & 2      & 2        & 3       & 4           & 3            & 2    & 3            & 1       & 432       & 20      \\
13 & 2      & 2        & 4       & 1           & 3            & 2    & 2            & 3       & 288       & 19      \\
14 & 2      & 5        & 2       & 3           & 4            & 3    & 3            & 4       & 4,320      & 38      \\
15 & 2      & 3        & 4       & 5           & 2            & 2    & 2            & 2       & 960       & 29      \\
16  & 3     & 6        & 5       & 2           & 7            & 3    & 4            & 13      & 65,520     & 309     \\
17  & 3     & 6        & 7       & 2           & 9            & 3    & 7            & 6       & 95,256     & 608     \\
18 & 3      & 6        & 5       & 6           & 9            & 4    & 8            & 3       & 155,520    & 461     \\
19 & 3      & 3        & 7       & 1           & 4            & 2    & 4            & 6       & 4,032      & 184     \\
20  & 3     & 6        & 2       & 6           & 5            & 5    & 7            & 9       & 113,400    & 440     \\
21 & 3      & 4        & 9       & 9           & 4            & 2    & 4            & 6       & 62,208     & 668     \\
22 & 3      & 8        & 5       & 2           & 7            & 3    & 5            & 8       & 67,200     & 367     \\
23 & 3      & 7        & 5       & 8           & 8            & 4    & 5            & 6       & 268,800    & 535     \\
24 & 3      & 5        & 5       & 7           & 8            & 4    & 9            & 2       & 100,800    & 638     \\
25 & 3      & 8        & 7       & 7           & 4            & 4    & 4            & 7       & 175,616    & 466     \\
26 & 4      & 8        & 9       & 2           & 10           & 3    & 8            & 13      & 449,280    & 1,733    \\
27 & 4      & 9        & 6       & 10          & 11           & 6    & 10           & 14      & 4,989,600   & 4,243    \\
28 & 4      & 6        & 10      & 9           & 11           & 4    & 12           & 7       & 1,995,840   & 5,916    \\
29 & 4      & 8        & 10      & 10          & 5            & 5    & 7            & 10      & 1,400,000   & 3,424    \\
30 & 4       & 9        & 7       & 9           & 12           & 6    & 10           & 10      & 4,082,400   & 4,272    \\
31 & 5      & 9        & 11      & 11          & 13           & 6    & 13           & 14      & 15,459,444  & 23,591 \\\hline  
\end{tabular}
\end{scriptsize}
\end{table*}

\begin{figure*}[!hbtp]
\centering
\includegraphics[scale=0.46]{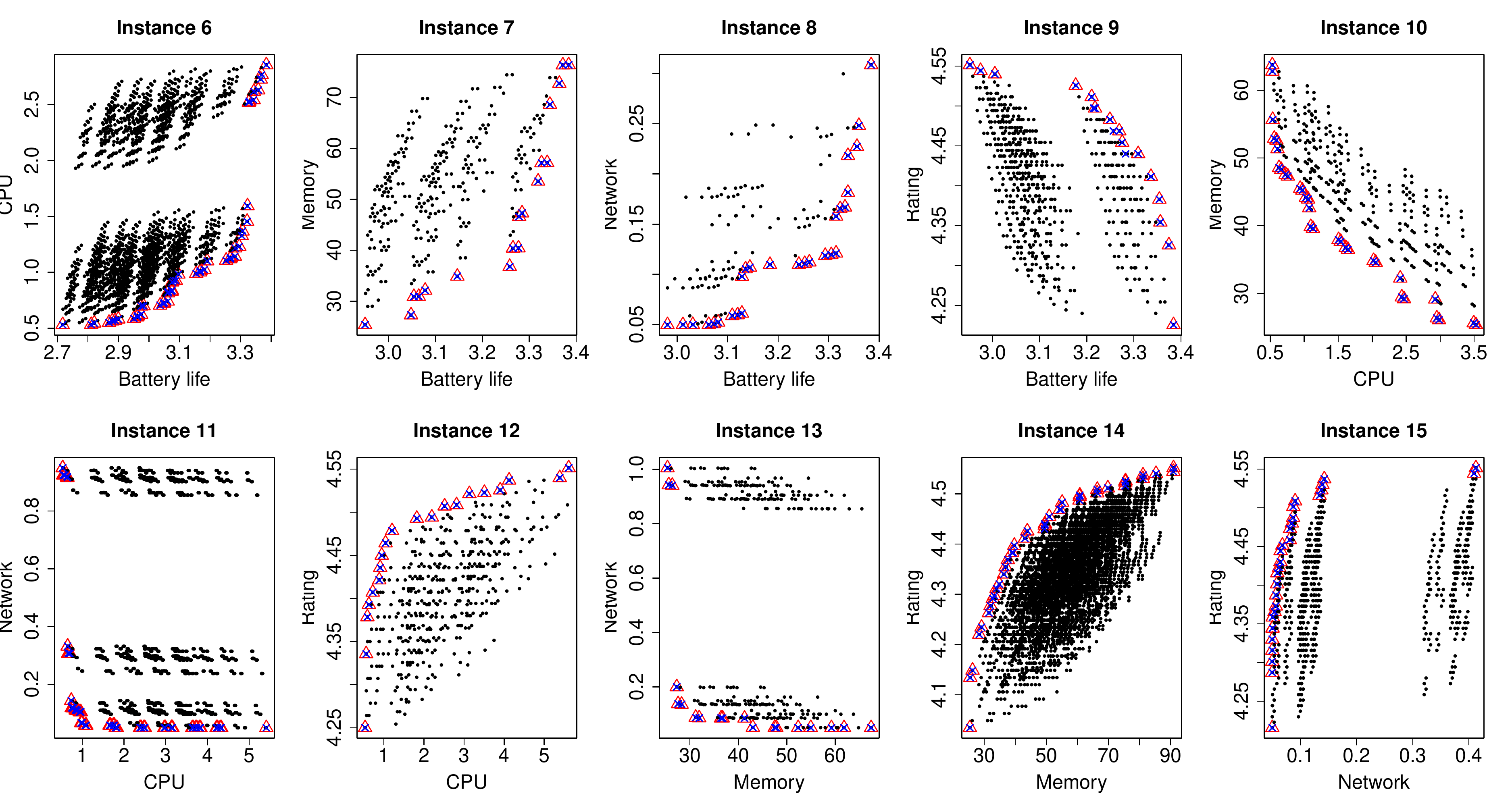}
\caption{Solutions of the bi-objective instances of the ASP for the Android case study. Symbol ($\bullet$) is used for all the possible solutions while Pareto optimal solutions are shown using the symbol ($\triangle$). Solutions found by APOA running \mbox{NSGA-II} are shown using the symbol ($\times$).}
\label{fig:PF-2D}
\end{figure*}

The exhaustive search took around 22 hours to solve all the instances of the ASP for the Android case study. Although \mbox{Instance 31} took 14 of those hours to solve. 
Using the EMO algorithm \mbox{NSGA-II}, all the instances are solved in less than two minutes. Pareto optimal solutions obtained by APOA after the exhaustive search and solutions found running \mbox{NSGA-II} for two dimensional instances of the ASP for the Android case study are shown in \mbox{Figure \ref{fig:PF-2D}} (instances with more than two objectives are not shown because the resulting plots are hard to read). In addition, non-optimal solutions after the search space reduction are also shown. This figure allows to compare the real Pareto optimal front obtained by the exhaustive search with respect to the approximation of the Pareto front generated by \mbox{NSGA-II}. As it is shown, solutions found by the exhaustive search and the EMO algorithm are overlapped which means that the latter is able to find the optimal solutions for these instances. This fact is expected, given that the search space is small in that cases. Concerning the rest of the instances, where the search space is bigger and more objective functions are involved, we check that \mbox{NSGA-II} is able to find solutions close to the optimal ones. From this we conclude that EMO algorithms as NSGA-II are a good alternative to the exhaustive search to solve the ASP. In addition to the previous, from \mbox{Figure \ref{fig:PF-2D}} we conclude, as we expected, that there exist a conflict between objectives because improving one metric implies to worsen the other. It is also interesting to note that there exist solution patterns in some instances. For example, for \mbox{Instance~13}, where memory and network usage are taken into account, two different clusters of solutions are clearly distinguished. A similar behavior exists for \mbox{Instance~6}, \mbox{Instance~9}, \mbox{Instance~11}, and \mbox{Instance~15}. We plan to study this phenomenon in a future work.

\subsection{Popular Versus Optimal Apps}
\cite{saborido_optimizing_2016} found that a good rating associated to an app in the marketplace does not warrant an efficient use of power and--or network.  In this subsection we check if this fact is also true for the four studied performance metrics.

We define the user solution as the
set of apps resulting after selecting the app with the maximum rating in each category. Considering the studied apps, the user solution is defined by the apps shown in \mbox{Table \ref{tab:userSolution}}. 
Let us define $\textbf{x}^{user}$ as the user
solution. Based on our measurements, \mbox{$power(\textbf{x}^{user}) = 2.81$}, \mbox{$CPU(\textbf{x}^{user}) = 8.92$}, \mbox{$memory(\textbf{x}^{user}) = 103.72$}, \mbox{$network(\textbf{x}^{user}) = 1.27$},
and \mbox{$rating(\textbf{x}^{user}) = 4.55$}, calculated using equations (\ref{eq:power}), (\ref{eq:cpu}), (\ref{eq:memory}), (\ref{eq:network}), and (\ref{eq:rating}). 

\begin{table}[H]
\centering
\caption{User solution -- Apps with maximum rating per category.}
\label{tab:userSolution}
\begin{scriptsize}
\begin{tabular}{ll}
\hline
Category &  App                                                  \\
\hline
     Browsers                     & mobi.mgeek.TunnyBrowser       \\
Cameras  &    com.roidapp.photogrid                               \\
Flash Lights  & goldenshorestechnologies.brightestflashlight.free  \\
News  & com.guardian                                              \\
Music Players                            & com.tbig.playerprotrial \\
Video Players                    &  video.player.audio.player.music\\
Weather                       &   com.handmark.expressweather    \\
\hline
\end{tabular}
\end{scriptsize}
\end{table}

\begin{figure}[H]
\centering
\includegraphics[scale=0.31]{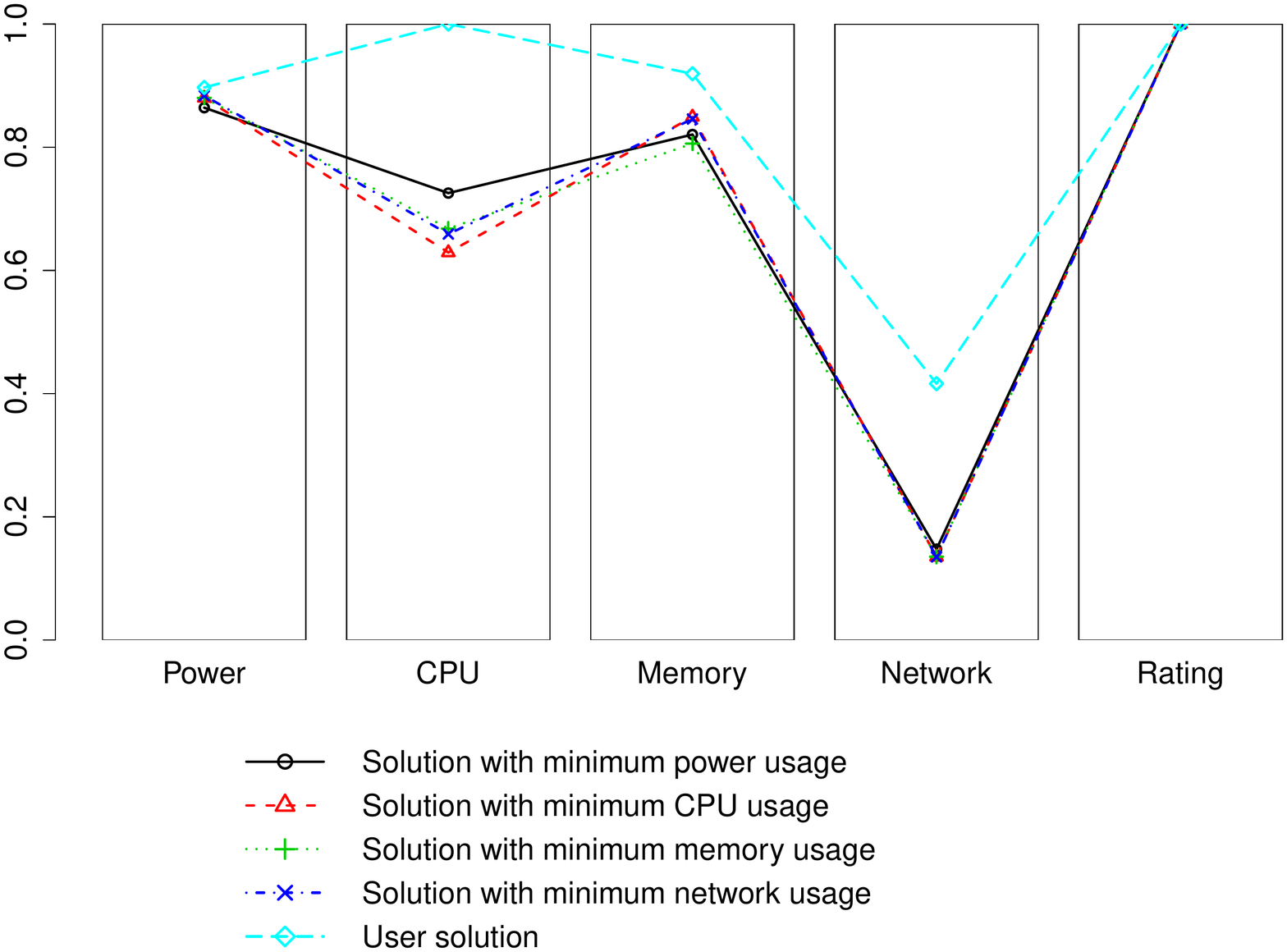}
\caption{ Comparison of four optimal solutions found by APOA for \mbox{Instance 31} of the ASP for the Android case study and the
solution chosen by the user. All of them with a rating value of 4.55.}
\label{fig:userVStool-sameRating}
\end{figure}

The user solution is compared to the optimal solutions found by APOA for \mbox{Instance 31} of the ASP for the \mbox{Android} case study, which optimize all the metrics simultaneously.
Out of 23,591 optimal solutions eight solutions have the same rating \comment{as} the user solution. From these eight solutions we select (1) the best one in terms of power usage ($\textbf{x}^{{min}_p}$), (2) the best one that
minimizes the CPU usage ($\textbf{x}^{{min}_{c}}$), (3) the best one in terms of memory usage ($\textbf{x}^{{min}_{m}}$), and (4) the best one that minimizes the network usage ($\textbf{x}^{{min}_n}$), and all of them are compared to the user solution in \mbox{Figure \ref{fig:userVStool-sameRating}}. In this chart, a bar is drawn
for each objective which are normalized to the interval $[0, 1]$ taking into account its maximum value considering the user solution and all the solutions in the Pareto optimal front obtained by APOA. A line is used to
represent each solution specifying the normalized value of each objective
function. 

As it is shown, APOA is able to find
better solutions in terms of performance metrics. The improvement in power, CPU, memory, and network usages is 3.66\%, 27.44\%, 10.73\%, and 64.57\% respectively, considering
$\textbf{x}^{{min}_{p}}$ versus {$\textbf{x}^{user}$}. This information is shown in \mbox{Table \ref{table:userVStool-sameRating}} for the four selected solutions. 
In all
the compared solutions the global rating is similar (4.55) but, as it is shown, performance metrics of
the user solution are worse than performance metrics of Pareto
optimal solutions found by APOA. Therefore the user solution is not optimal. We corroborate our prior findings that a good rating associated to an app in the Google Play marketplace does not warrant an efficient use of power and--or network usages, but we extend this fact to CPU and memory usages.

\begin{table}[!htbp]
\centering
\caption{
Improvement of optimal solutions versus user solution.
Depicted are improvements in \% for each metric of optimal solutions found by APOA for \mbox{Instance 31} for the ASP from the Android case study are used instead of the solution chosen by the user. All of these solutions have a rating value of 4.55.}
\label{table:userVStool-sameRating}
\begin{scriptsize}
\begin{tabular}{cccccc}
\hline
Solution           & Power & CPU   & Memory & Network & Rating \\
\hline
$\textbf{x}^{{min}_p}$   & \textbf{3.66}  & 27.44 & 10.73  & 64.57   & 0.00   \\
$\textbf{x}^{{min}_c}$     & 1.90  & \textbf{37.09} & 7.63   & 67.57   & 0.00   \\
$\textbf{x}^{{min}_m}$  & 1.88  & 33.20 & \textbf{12.33}  & 67.54   & 0.00   \\
$\textbf{x}^{{min}_n}$ & 1.48  & 34.10 & 8.00   & \textbf{67.57}   & 0.00  \\
\hline
\end{tabular}
\end{scriptsize}
\end{table}

\begin{figure}[H]
\centering
\includegraphics[scale=0.31]{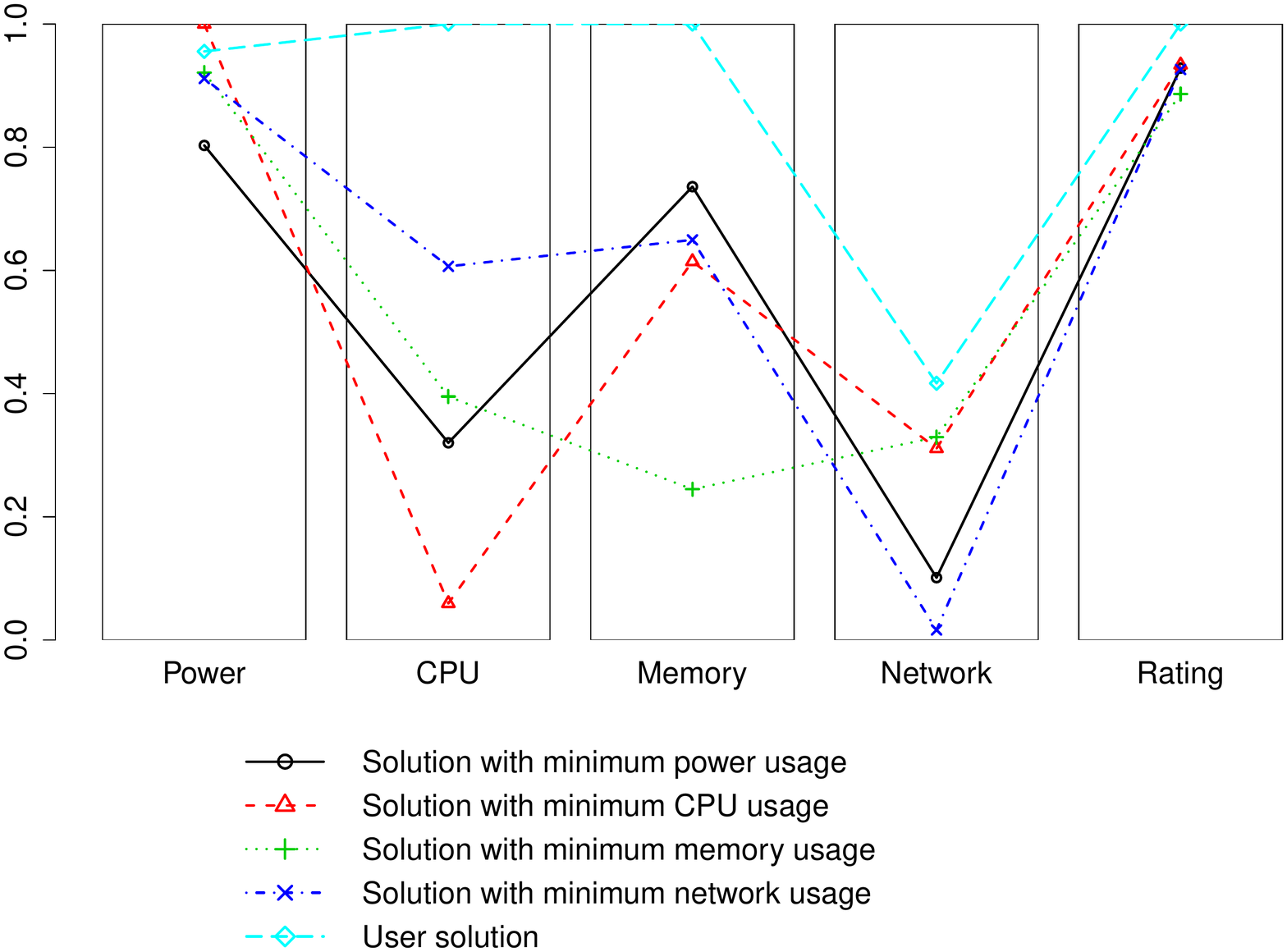}
\caption{Comparison of four optimal solutions found by APOA for \mbox{Instance 26} of the ASP for the Android case study and the solution chosen by the user. Sacrificing the user rating performance metrics greatly improves performance measures. 
}
\label{fig:userVStool}
\end{figure}

Now we shall investigate preferring  performance metrics more than user ratings. Effectively we study the improvement in performance when the user rating is sacrificed. In order to carry out this study we analyze the Pareto optimal solutions found by APOA for \mbox{Instance 26} of the ASP for the Android case study, which optimize performance metrics (but not the rating). Out of 1,733 optimal solutions we select the following four solutions: (1) the best one in terms of power usage ($\textbf{x'}^{({{min}_p})}$), (2) the best one that
minimizes the CPU usage ($\textbf{x'}^{{min}_{c}}$), (3) the best one in terms of memory usage ($\textbf{x'}^{{min}_{m}}$), and (4) the best one that minimizes the network usage ($\textbf{x'}^{({{min}_n})}$). 
In \mbox{Figure \ref{fig:userVStool}} we compare these solutions versus the user solution in terms of performance and rating. 

Regarding our data, when $\textbf{x'}^{({{min}_p})}$ is compared to
$\textbf{x}^{user}$, the rating is sacrificed by 7.16\%, but the improvement in
power, CPU, memory, and network usages is 15.98\%, 67.99\%, 26.40\%, and 75.73\%,
respectively. This information is shown in \mbox{Table \ref{table:userVStool}} for each of the selected solutions. We conclude that, sacrificing the rating in a small magnitude, performance metrics are improved in a big proportion.

\begin{table}[!htbp]
\centering
\caption{Improvement (in \%) for each metric when optimal solutions found by APOA for \mbox{Instance 26} of the ASP for the Android case study are used instead of the solution chosen by the user. Negative values indicate that a solution found by APOA worsens with respect to the user solution.}
\label{table:userVStool}
\begin{scriptsize}
\begin{tabular}{cccccc}
\hline
Solution           & Power & CPU   & Memory & Network & Rating \\
\hline
$\textbf{x'}^{({{min}_p})}$   & \textbf{15.98} & 67.99 & 26.40  & 75.73   & -7.16  \\
$\textbf{x'}^{({{min}_c})}$     & -4.44 & \textbf{94.06} & 38.55  & 25.39   & -6.62  \\
$\textbf{x'}^{({{min}_m})}$  & 3.60  & 60.47 & \textbf{75.53}  & 20.97   & -11.36 \\
$\textbf{x'}^{({{min}_n})}$ & 4.60  & 39.34 & 35.04  & \textbf{96.09}   & -7.38  \\
\hline
\end{tabular}
\end{scriptsize}
\end{table}

\subsection{APOA and Contexts of Use}
In this subsection we use APOA in the contexts of use described previously in \mbox{Section \ref{sec:contexts}} and, for each of them, we show the Pareto optimal solutions and the existing trade-off associated to each solution and performance metric.

\subsubsection*{Travel Abroad}
This context of use considers users who travel abroad (for working or holidays). In that case we consider battery life and network usage the most important metrics. First, because in this context likely the time between consecutive charges uses to be longer and, second, because data roaming is usually expensive. This context of use corresponds to \mbox{Instance 8} of the ASP, whose Pareto optimal front was previously presented in \mbox{Figure \ref{fig:PF-2D}}. Out of 120 possible solutions 27 (22.50\%) are Pareto optimal in terms of battery life and network usage. \mbox{Table \ref{table:solutionsInstance8}} shows these solutions. The second and third columns show the objective values of these performance metrics. Columns fourth and fifth show the trade-off associated to each solution and performance metric. 

\begin{table}[!htbp]
\centering
\caption{Pareto optimal solutions and their associated trade-off for the \textit{Travel Abroad} context of use. Objective values for battery life and network usage are expressed in hours and MB, respectively.}
\label{table:solutionsInstance8}
\begin{scriptsize}
\begin{tabular}{ccccc}
\hline
         & \multicolumn{2}{c}{Objective values}       & \multicolumn{2}{c}{Trade-off (in \%)} \\
Solution & Battery & Network  & Battery         & Network        \\
\hline
1        & 3.38                     & 0.31            & 0.00                 & 83.88          \\
2        & 3.36                     & 0.25            & 0.71                 & 64.30          \\
3        & 3.36                     & 0.23            & 0.83                 & 57.54          \\
4        & 3.34                     & 0.22            & 1.36                 & 54.60          \\
5        & 3.34                     & 0.18            & 1.36                 & 42.70          \\
6        & 3.33                     & 0.17            & 1.53                 & 37.96          \\
7        & 3.32                     & 0.17            & 1.81                 & 37.45          \\
8        & 3.31                     & 0.16            & 2.05                 & 35.02          \\
9        & 3.31                     & 0.12            & 2.05                 & 23.12          \\
10       & 3.31                     & 0.12            & 2.33                 & 22.61          \\
11       & 3.29                     & 0.12            & 2.69                 & 22.35          \\
12       & 3.26                     & 0.11            & 3.61                 & 20.24          \\
13       & 3.25                     & 0.11            & 3.89                 & 19.72          \\
14       & 3.24                     & 0.11            & 4.23                 & 19.47          \\
15       & 3.18                     & 0.11            & 5.92                 & 19.42          \\
16       & 3.14                     & 0.11            & 7.10                 & 18.54          \\
17       & 3.14                     & 0.11            & 7.35                 & 18.03          \\
18       & 3.13                     & 0.10            & 7.56                 & 15.60          \\
19       & 3.13                     & 0.06            & 7.56                 & 3.70           \\
20       & 3.12                     & 0.06            & 7.81                 & 3.18           \\
21       & 3.11                     & 0.06            & 8.13                 & 2.92           \\
22       & 3.08                     & 0.05            & 8.96                 & 0.82           \\
23       & 3.07                     & 0.05            & 9.20                 & 0.30           \\
24       & 3.06                     & 0.05            & 9.50                 & 0.04           \\
25       & 3.03                     & 0.05            & 10.44                & 0.04           \\
26       & 3.01                     & 0.05            & 11.02                & 0.00           \\
27       & 2.98                     & 0.05            & 11.92                & 0.00          \\
\hline
\end{tabular}
\end{scriptsize}
\end{table}

\begin{figure}[!htbp]
\centering
\includegraphics[scale=0.7]{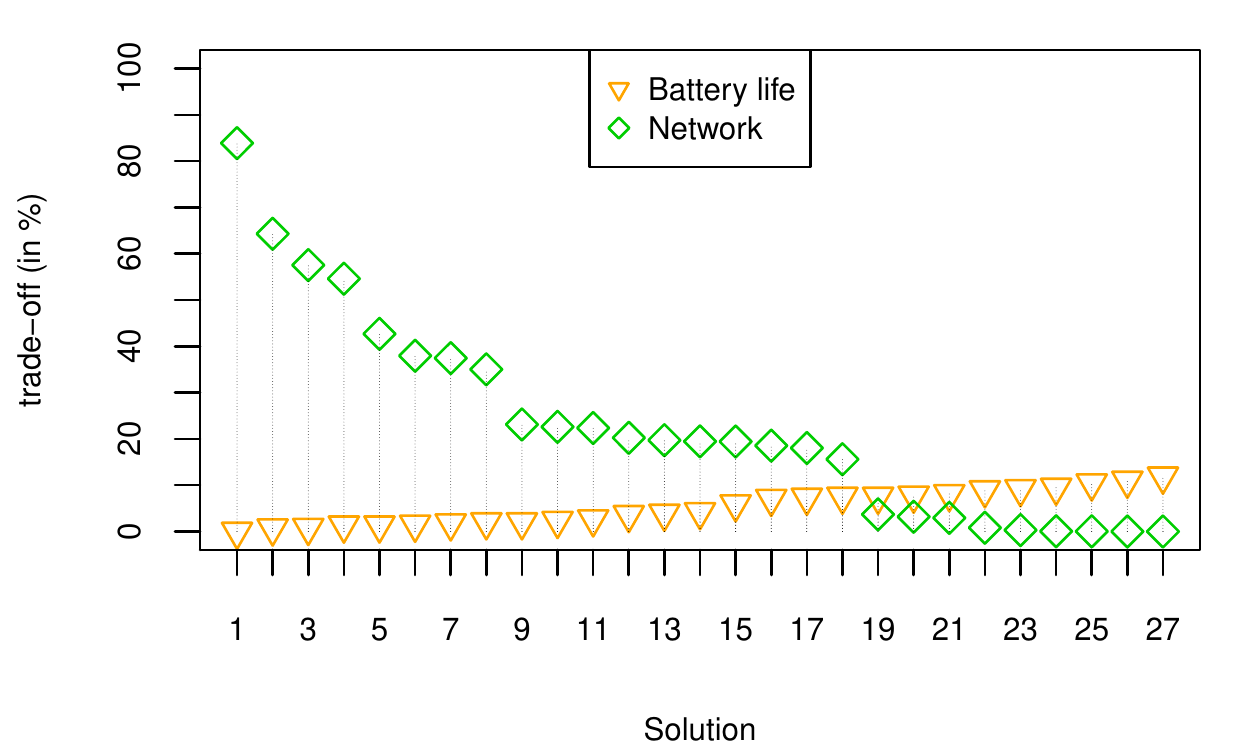}
\caption{Trade-off (in \%) for battery life and network usage of each solution for the \textit{Travel Abroad} context of use.}
\label{fig:tradeoffForInstance8}
\end{figure}

\mbox{Figure \ref{fig:tradeoffForInstance8}} shows using a bars plot the trade-off of each solution for battery life and network usage. This plot and the previous table, used together, are useful for the user to visualize and compare Pareto optimal solutions. If the user prefers battery life to network usage, \mbox{Solution 1} would be chosen. In that case network usage is increased 0.26 MB (83.88\%) with respect to \mbox{Solution 27}, which has the lowest network usage. On the contrary, if network usage is preferred, \mbox{Solution 27} would be chosen decreasing battery life 24 minutes (11.92\%) with respect to \mbox{Solution 1}. If both performance metrics are equally important, \mbox{Solution 19} could be chosen as the preferred one because the trade-off is almost similar for both objective functions. In that case battery life is decreased 15 minutes (7.56\%) respect to \mbox{Solution 1} and network usage is increased 0.01 MB (3.70\%) with respect to \mbox{Solution 27}.

\subsubsection*{Old Devices}
\begin{figure}[!htbp]
\centering
\includegraphics[scale=0.7]{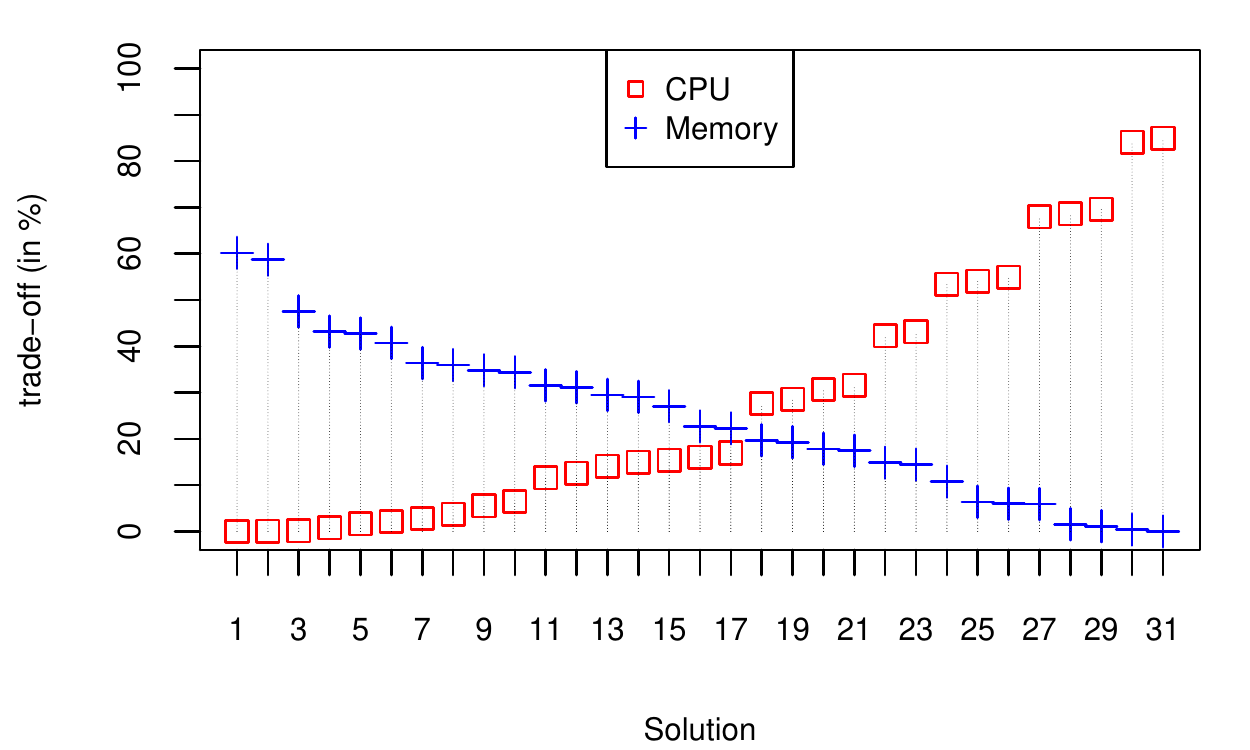}
\caption{Trade-off (in \%) for CPU and memory usages of each solution for the \textit{Old Devices} context of use. 
}
\label{fig:tradeoffForInstance10}
\end{figure}

This context of use considers users who own old devices which are limited in terms of CPU and RAM memory. In that case we consider CPU and memory usages as the most important performance metrics. This context of use corresponds to \mbox{Instance 10} of the ASP, whose Pareto optimal front was previously presented in \mbox{Figure \ref{fig:PF-2D}}. Over 320 existing solutions 31 (9.69\%) are Pareto optimal in terms of CPU and memory usages. \mbox{Table \ref{table:solutionsInstance10}} shows these solutions, their metric values, and the trade-off associated to each solution and performance metric. \mbox{Figure \ref{fig:tradeoffForInstance10}} shows the trade-off of each solution for CPU and memory usages. If the user prefers CPU to memory usage, \mbox{Solution 1} would be chosen. In that case memory usage is increased 38.36 MB (60.19\%) with respect to \mbox{Solution 31}, which uses less memory. On the contrary, if memory usage is preferred, \mbox{Solution 31} would be chosen increasing CPU usage 3.00\% (trade-off of 84.96\%)
with respect to \mbox{Solution 1}. If CPU and memory usages are equally important, \mbox{Solution 17} could be chosen. However, if memory usage is a little more important than CPU usage, Solutions from 18 to 31 could be preferred.

\begin{table}[!htbp]
\centering
\caption{Pareto optimal solutions and their associated trade-off for the \textit{Old Devices} context of use. Objective values for CPU and memory usages are expressed in percentage and MB, respectively.}
\label{table:solutionsInstance10}
\begin{scriptsize}
\begin{tabular}{ccccc}
\hline
         & \multicolumn{2}{c}{Objective values} & \multicolumn{2}{c}{Trade-off (in \%)} \\
Solution & CPU     & Memory     & CPU               & Memory            \\
\hline
1        & 0.53            & 63.74              & 0.00              & 60.19             \\
2        & 0.53            & 62.79              & 0.04              & 58.70             \\
3        & 0.54            & 55.67              & 0.15              & 47.53             \\
4        & 0.56            & 52.89              & 0.81              & 43.17             \\
5        & 0.59            & 52.63              & 1.71              & 42.75             \\
6        & 0.61            & 51.34              & 2.16              & 40.73             \\
7        & 0.63            & 48.56              & 2.81              & 36.37             \\
8        & 0.66            & 48.29              & 3.72              & 35.95             \\
9        & 0.73            & 47.57              & 5.58              & 34.82             \\
10       & 0.76            & 47.30              & 6.48              & 34.40             \\
11       & 0.94            & 45.50              & 11.63             & 31.57             \\
12       & 0.97            & 45.23              & 12.54             & 31.15             \\
13       & 1.03            & 44.18              & 14.05             & 29.50             \\
14       & 1.06            & 43.91              & 14.95             & 29.08             \\
15       & 1.07            & 42.63              & 15.39             & 27.06             \\
16       & 1.10            & 39.85              & 16.05             & 22.70             \\
17       & 1.13            & 39.58              & 16.96             & 22.28             \\
18       & 1.51            & 37.93              & 27.65             & 19.70             \\
19       & 1.54            & 37.67              & 28.56             & 19.28             \\
20       & 1.61            & 36.75              & 30.66             & 17.85             \\
21       & 1.64            & 36.49              & 31.57             & 17.43             \\
22       & 2.02            & 34.84              & 42.26             & 14.85             \\
23       & 2.05            & 34.57              & 43.16             & 14.43             \\
24       & 2.41            & 32.25              & 53.39             & 10.77             \\
25       & 2.44            & 29.46              & 54.05             & 6.41              \\
26       & 2.47            & 29.20              & 54.95             & 5.99              \\
27       & 2.93            & 29.15              & 68.00             & 5.92              \\
28       & 2.95            & 26.37              & 68.66             & 1.56              \\
29       & 2.98            & 26.10              & 69.56             & 1.14              \\
30       & 3.49            & 25.64              & 84.06             & 0.42              \\
31       & 3.53            & 25.38              & 84.96             & 0.00         \\
\hline    
\end{tabular}
\end{scriptsize}
\end{table}

\subsubsection*{Driving}
This context includes users who use their mobile devices as GPS navigator. As we mentioned in \mbox{Section \ref{sec:contexts}}, we divide this context of use in two different subcontexts.

\begin{enumerate}
\item{Phone is not plugged to the car charger. Here we consider energy consumption as the most important performance metric. This context of use corresponds to \mbox{Instance 1} of the ASP. There is only one optimal combination of apps minimizing power usage, which has associated a battery life of 3.38 hours.}

\item{Phone is plugged to the car charger. In this case we consider that the most important performance metrics are the network usage (to minimize data plan usage), and CPU and memory usages (to avoid lags in navigation due to the phone's slow down). This context of use corresponds to \mbox{Instance 22} of the ASP. As it is shown in \mbox{Table \ref{tab:solutions}}, out of 67,200 possible solutions 367 (0.55\%) are Pareto optimal in terms of CPU, memory, and network usages. Because of space limitation these Pareto optimal solutions cannot be enumerated in the paper but 
\mbox{Figure \ref{fig:tradeoffForInstance10}} shows the trade-off of each solution for CPU, memory, and network usage. If CPU and memory usages are equally important but a trade-off for network usage lower than 10.00\% is preferred, any solution from 1 to 196 could be chosen.


\begin{figure*}[!htbp]
\centering
\includegraphics[scale=0.7]{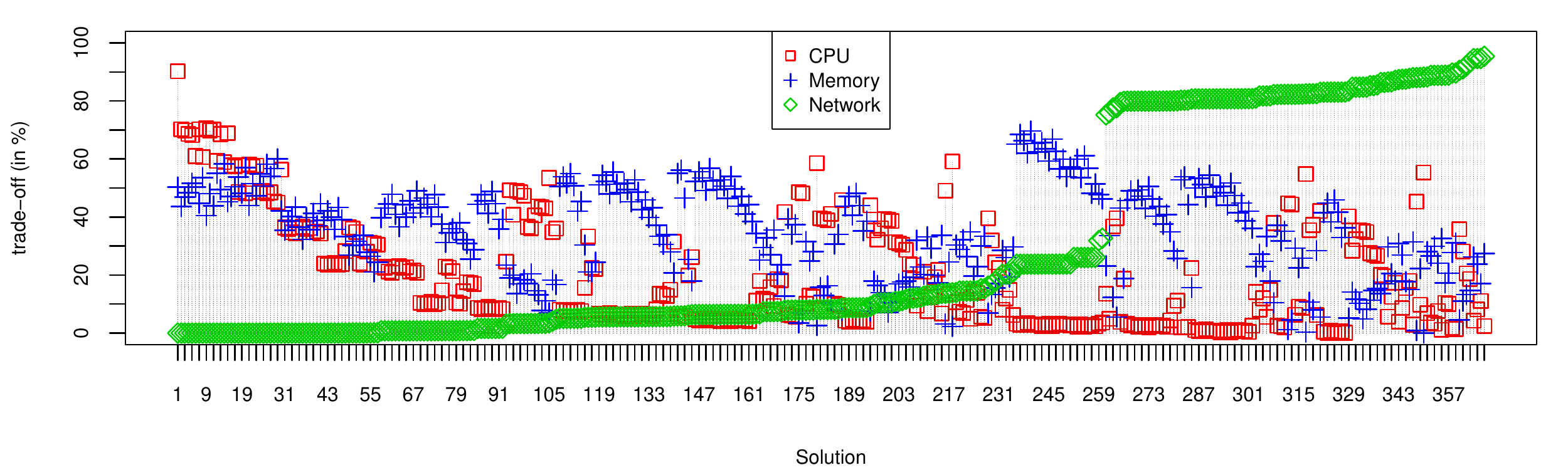}
\caption{Trade-off (in \%) for CPU, memory, and network usages of each solution for the \textit{Driving} context of use when the phone is plugged to the car charger.}
\label{fig:tradeoffForInstance22}
\end{figure*}
}
\end{enumerate}



\noindent
Depending on the context of use some metrics could be more important than others and this fact could affect users' preferences. As we showed, even if the number of Pareto optimal solutions for a given context is much less than the number of initial possible solutions, still users should decide the preferred one according to their preferences. For this reason, in addition to the list of Pareto optimal solutions, information about the existing trade-off between solutions and metrics should be given to the user.

\subsection{Assisting Developers in Comparing Apps}

We showed that APOA can help users to make efficient choices about apps in terms of performance metrics. Performance metrics are not only important for mobile device users but also for developers. For example, given a new app or a new version of an existing one, knowing how close or far in terms of performance metrics is the new app to the others in the same category. In this subsection we show how collected measures can be used to assist developers in comparing new apps with respect to their competitors in the marketplace.

Collected data are used to define the reference values shown in Table \ref{table:forDevelopers}. Here, regarding the apps in each category, the optimal, median, and worst values for each performance metric are shown. Let us suppose that a new Android browser app is developed. The scenario associated to the browsers category is played and performance metrics are collected. Let us consider that the associated power usage (battery life), CPU, memory, and data usage are 3.10W (2.57 hours), 9.00\%, 65.00MB, and 0.40MB, respectively, for this new app. Using the reference values given in Table \ref{table:forDevelopers} we can know that power usage (battery life) associated to the app is worse than battery life associated to most of the apps in that category. 
Concerning CPU usage, the new app is the worst but it is the best in terms of memory usage. In addition, network usage of the new app does not improve the best existing value, but it is better than the median.

\begin{table*}[!htbp]
\centering
\caption{Reference values for performance metrics of Android app categories.}
\label{table:forDevelopers}
\begin{scriptsize}
\begin{tabular}{lcccccccccccccc}
\hline
\multirow{2}{*}{Category} & \multicolumn{3}{c}{Battery life (in hours)} & \multicolumn{3}{c}{CPU (in \%)}  & \multicolumn{3}{c}{Memory (in MB)} & \multicolumn{3}{c}{Network (in MB)} 
\\
                          & Optimal    & Median    & Worst   & Optimal & Median & Worst & Optimal  & Median  & Worst & Optimal  & Median  & Worst 
                          \\
\hline
Browsers                  & 2.79       & 2.65      & 2.46    & 0.95    & 2.76   & 7.92  & 0.07     & 0.12    & 0.17  & 0.24     & 0.62    & 1.38   \\
Cameras                   & 2.69       & 2.24      & 1.90    & 0.00    & 1.78   & 12.59 & 0.01     & 0.09    & 0.15  & 0.00     & 0.02    & 3.44   \\
Flash Lights               &    11.90    & 8.82      & 5.39   & 0.00    & 4.03   & 13.93 & 0.01     & 0.07    & 0.11  & 0.00     & 0.01    & 1.21  \\
Music Players             &    4.64    & 2.74      & 2.04    & 0.00    & 5.07   & 12.95 & 0.02     & 0.08    & 0.10  & 0.00     & 0.00    & 0.42  \\
News                      &    2.87    & 2.61      & 2.18    & 1.12    & 10.46  & 17.48 & 0.05     & 0.15    & 0.27  & 0.08     & 1.36    & 9.16   \\
Video Players              &   2.75     & 2.37      & 1.79    & 0.11    & 9.34   & 13.90 & 0.01     & 0.09    & 0.12  & 0.00     & 0.00    & 0.39   \\
Weather                   &  3.06      & 2.63      & 1.91    & 1.53    & 6.90   & 19.44 & 0.01     & 0.10    & 0.18  & 0.02     & 0.43    & 14.00  \\
\hline
\end{tabular}
\end{scriptsize}
\end{table*}

\begin{figure}[!htb] 
  \begin{subfigure}[b]{0.5\linewidth}
    \centering
    \includegraphics[width=1\linewidth]{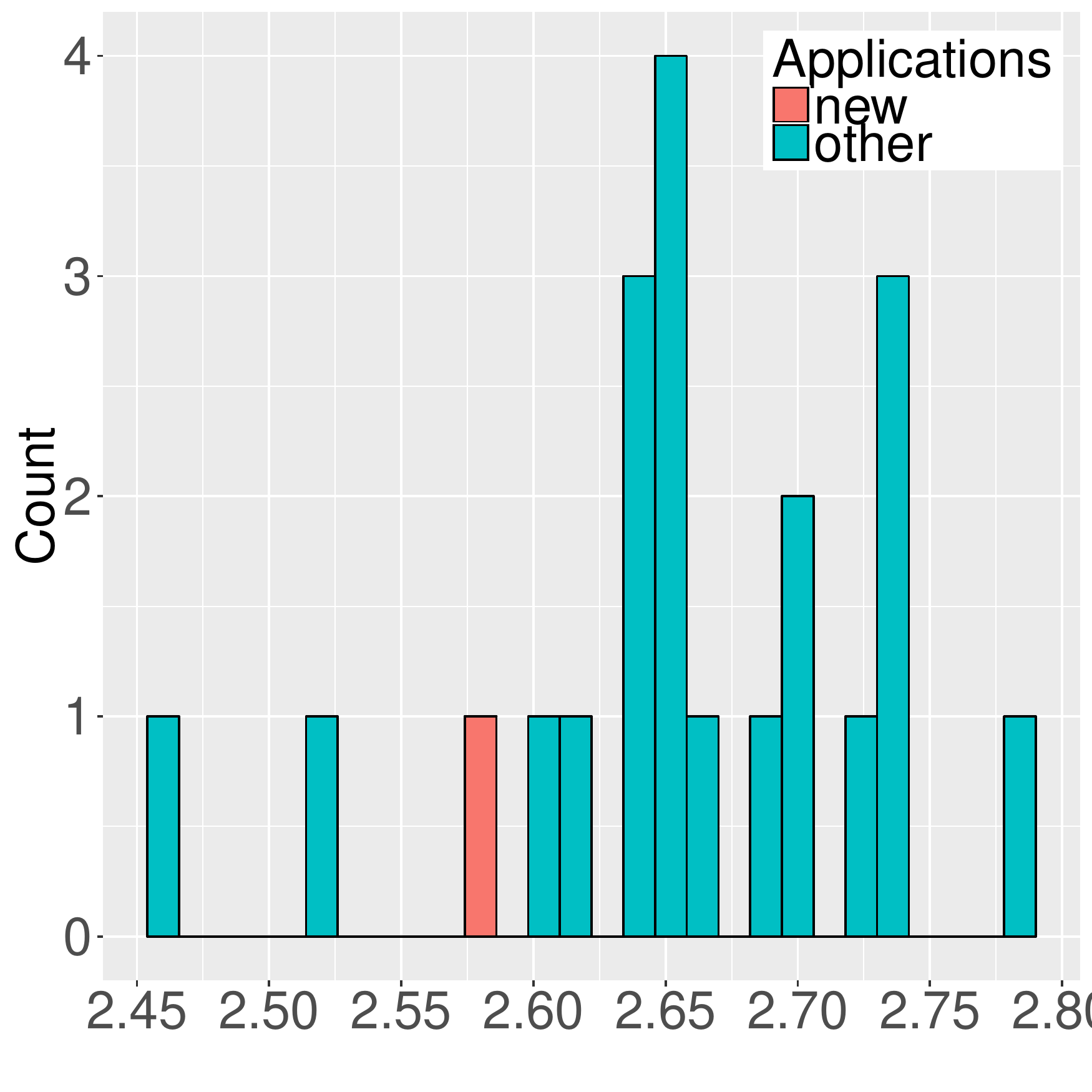} 
    \caption{Battery life (in hours)} 
    \label{fig:histogramBattery} 
    \vspace{4ex}
  \end{subfigure}
  \begin{subfigure}[b]{0.5\linewidth}
    \centering
    \includegraphics[width=1\linewidth]{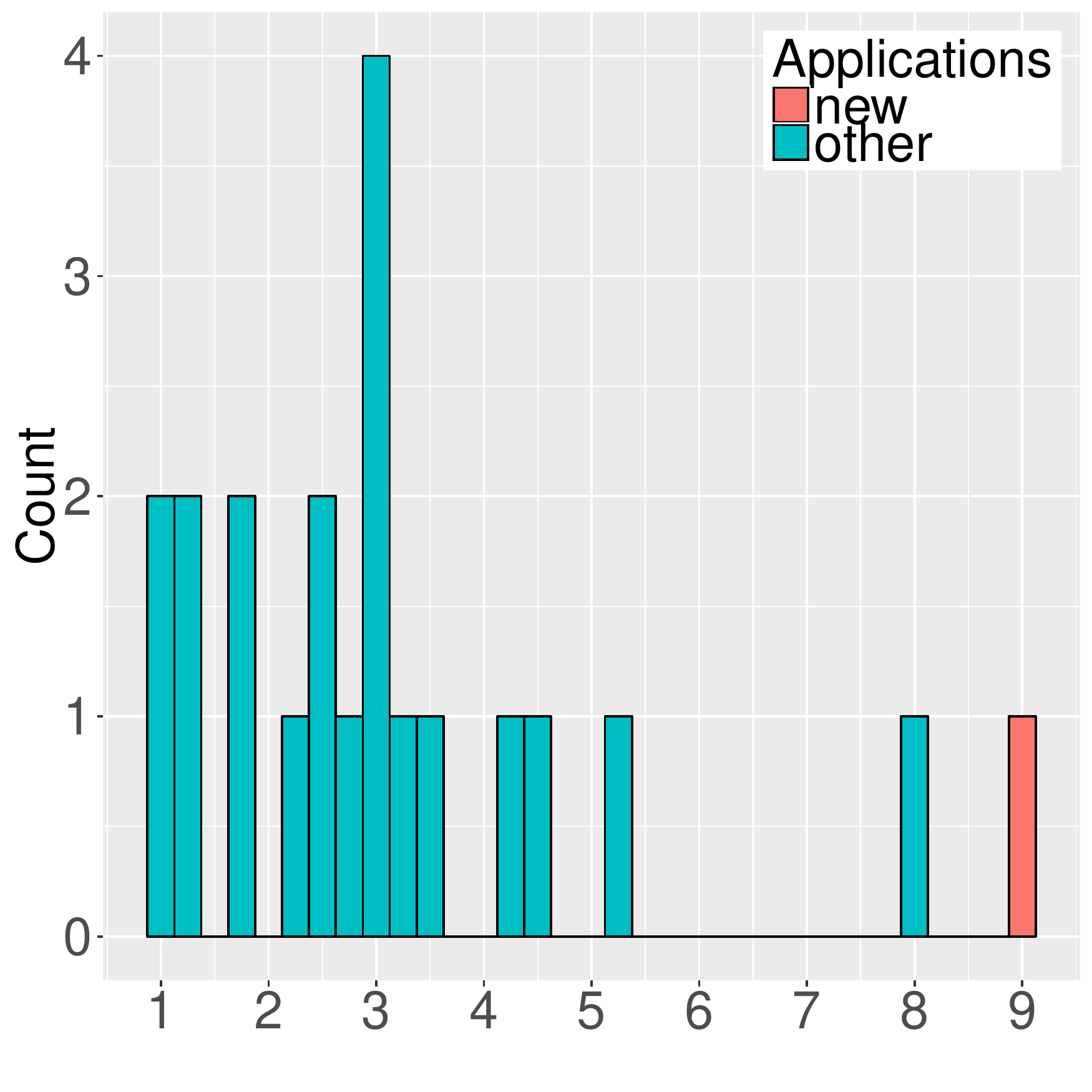} 
    \caption{CPU usage (in \%)} 
    \label{fig:histogramCPU} 
    \vspace{4ex}
  \end{subfigure} 
  \begin{subfigure}[b]{0.5\linewidth}
    \centering
    \includegraphics[width=1\linewidth]{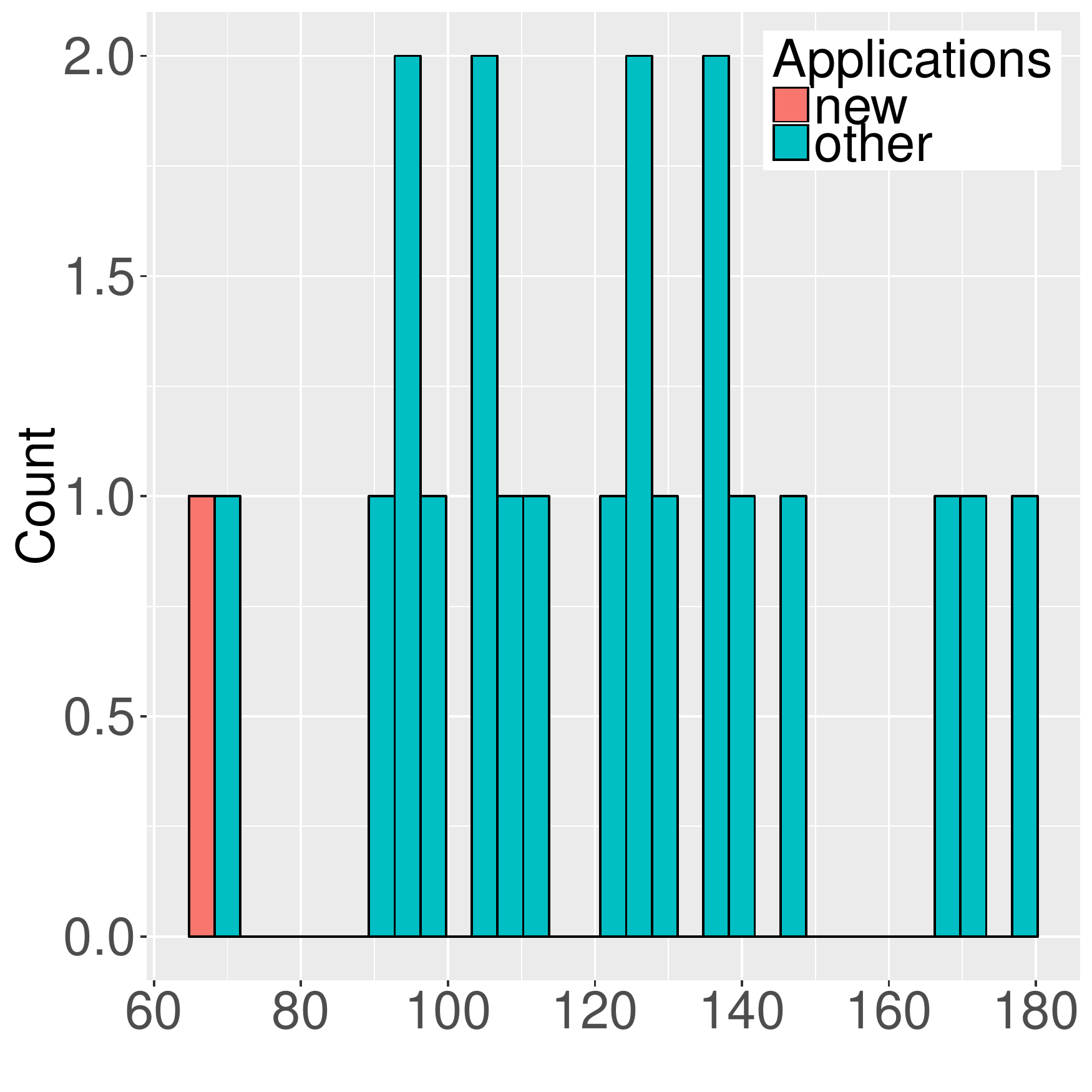} 
    \caption{Memory usage (in MB)} 
    \label{fig:histogramMemory} 
  \end{subfigure}
  \begin{subfigure}[b]{0.5\linewidth}
    \centering
    \includegraphics[width=1\linewidth]{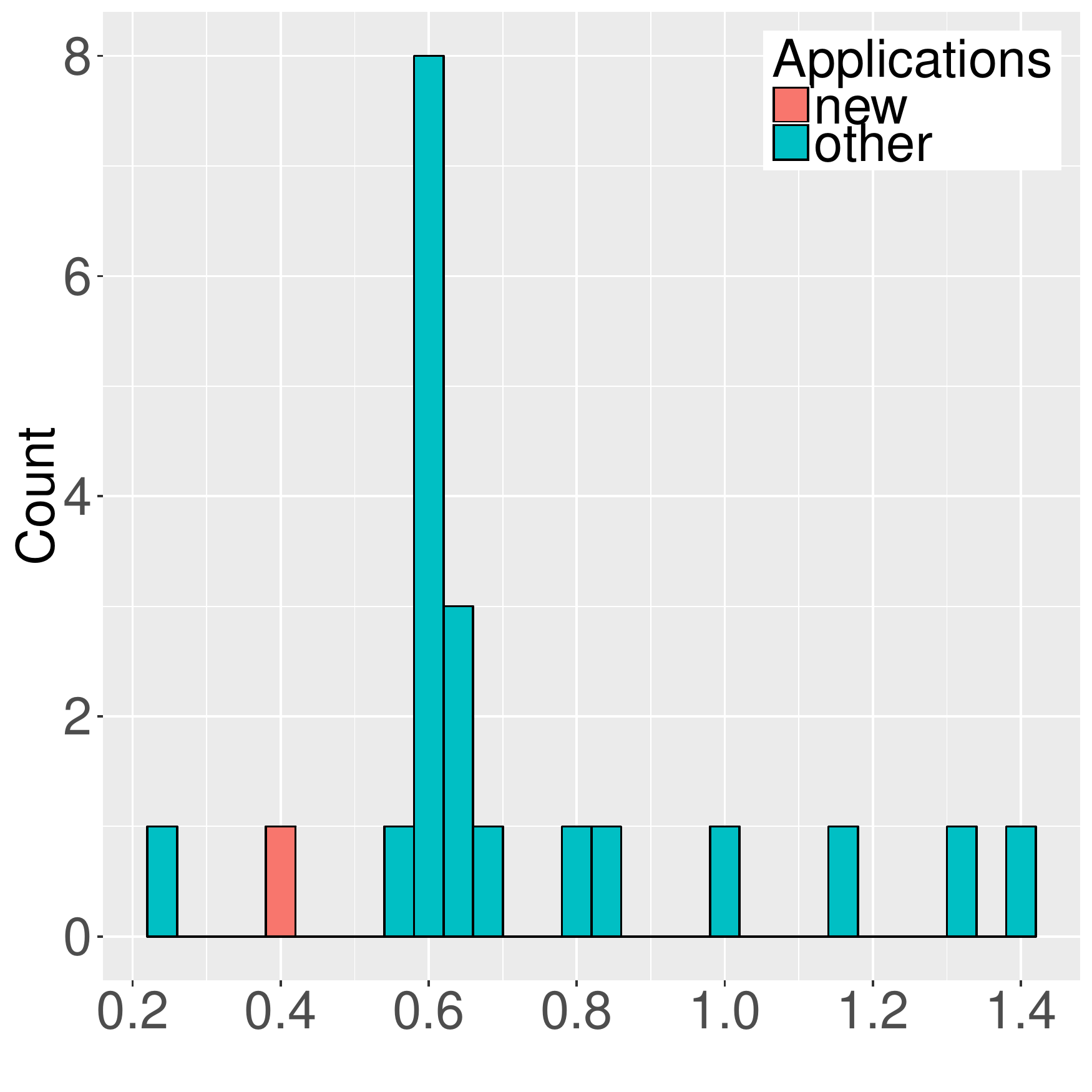} 
    \caption{Network usage (in MB)} 
    \label{fig:histogramNetwork} 
  \end{subfigure} 
  \caption{Distribution of performance metrics for browser apps from our case study. It allows to know how a new app is positioned in terms of performance metrics with respect to other apps. 
}
  \label{fig:histograms} 
\end{figure}

Using histograms, developers can visualize, in a graphical way, the distribution of performance metrics of apps belonging to the same category. It allows developers to know how their new app is positioned with respect to the others. Figure \ref{fig:histograms} shows the histograms of apps in the browsers category for each performance metric (where the red bar represents the new app). 

Analyzing the histograms, we confirm that the new app is the third worst in terms of power usage (battery life), the worst in terms of CPU usage, but the best in terms of memory usage. Considering network usage, we conclude that the new app is the second best app in the category with respect to data usage.

\section{Threats to Validity}
\label{sec:threats}
Threats to internal validity concern factors, internal to our study, that could have influenced our results. For our case study we computed performance metrics using well-known approaches and scenarios. In addition, we replicated several times our measures to ensure statistical validity.

Threats to construct validity concern relationship between theory and observation and the extent to which the measures represent real values. We used a Nexus 4 phone, the same model used in previous studies~(\cite{di_nucci_software-based_2017,huang_defdroid:_2016,linares-vasquez_mining_2014,saborido_optimizing_2016,sahin_how_2014,sahin_benchmarks_2016}). Our measurement apparatus had a higher number of sampling bits \comment{than} previous studies and our sampling frequency was one order of magnitude higher than past studies. Overall, our measures were more precise or at least as precise as those in previous studies. Network, CPU and memory usages were collected using the tool \texttt{tcpdump} and the commands \texttt{top} and \texttt{dumpsys} on the phone, respectively, which may have introduced extra energy consumption. 
In order to avoid any impact of other metrics' measurements on it, we collected power usage individually.
\comment{Another threat to construct validity is the scenarios and default application settings. We only explore the default application settings but there could be drastically different performance if these settings were to change. Unfortunately we lack statistics about this behavior and its prevalence, thus it remains a threat that faces this study.}

Threats to conclusion validity concern the relationship between experimentation and outcome. Further validations on different marketplaces, larger set of apps, and--or different phones is desirable to make our findings more generic. 

Threats to external validity concern the generalization of our findings. The case study was limited to a reduced number of popular Android apps in the Android marketplace. More experiments over larger set of apps must be carried out to evaluate APOA in a more realistic way using tens of categories and million apps. In that case the exhaustive search could not be feasible and \mbox{NSGA-II} could have some limitations to solve instances where the number of objectives is greater than three (\cite{Khare2003,Ishibuchi:2008}). For this reason more algorithms should be studied. Furthermore apps evolve and so does their performance, this kind of work should extend across multiple versions (\cite{hindle_green_2012}). 

Another concern in this research is whether or not the scenarios are representative of the overall app behavior. We are aware that \comment{(i) different features have different performance characteristics and (ii) there is no a clear relationship between the rating of apps and their number of features.} \comment{Thus,} more experiments considering different scenarios \comment{and features} is required. \comment{Although our goal was to show APOA feasibility and support the evidence of its usefulness, we made an effort to define complete scenarios simulating users behaviour exercising more common app features. Thus, for example, we got the forecast for two different cities for apps in the Weather category.}

In order to allow \comment{the replication of} this study and reduce the threats to validity, all the content used is available in our replication package\footnote{http://www.ptidej.net/downloads/replications/APOA/}. It is important to notice that the same  model of phone and version of Android operating system should be used to replicate the case study.

\section{Discussion about Performance Measures}
\label{sec:discussion}
Energy impact of apps running on mobile devices can be measured using hardware based approaches, as the methodology explained in this paper, or using software based approaches, as PETrA (\cite{di_nucci_software-based_2017}) or Xcode\footnote{https://developer.apple.com/xcode/} 
for Android and iOS platforms, respectively. However, existing
services such as Green Miner \mbox{(\cite{hindle_greenminer:_2014})} could also be used. Green Miner is a dedicated hardware mining software repositories testbed. It physically measures the energy consumption of mobile phones and automates the reporting of measurements as a web service.

CPU, memory, and network usages can be measured using software based approaches. The Android tools used in our case study can be used to collect information about these metrics in Android platforms. For iOS\comment{,} Xcode instruments 
can be used to automate the collection of these performance metrics in Apple platforms.

\comment{In the Android and Apple app stores there are million apps. Thus, performance metrics collection for all apps is infeasible in practice using the solution described in the APOA case study. Although APOA is independent to the measurement process, hybrid static and dynamic analysis techniques could be applied to estimate performance metrics of apps in a marketplace. Based on this idea, \mbox{\cite{behrouz_ecodroid:_2015}} proposed EcoDroid to estimate the energy cost of Android apps taking into account their API usage. A similar approach could be applied to estimate CPU, memory, and network usages of apps. In addition to this, Google has recently announced, at the I/O 2017 conference, Android Vitals\footnote{https://developer.android.com/distribute/best-practices/develop/android-vitals.html}. Vitals identifies different issues in Android apps from Android devices whose users have opted-in to automatically share usage and diagnostics data. The Play Console aggregates this information and displays metrics about stability, rendering time, and battery usage, in a specific dashboard. APOA could also be implemented in the Android marketplace getting different metrics of apps from Vitals.}

\comment{Nonetheless, performance metrics measurements could be done as part of a certification process by a third party who is potentially paid to test and certify apps. This puts the onus on a business to engage in this practice to provide badges or approvals to apps that might garner consumer trust---much like consumer reports.
}



\section{Related Work}
\label{sec:relatedWork}
\label{sec:relatedWork}
\comment{There is a growing body of work on analyzing and optimizing the performance of mobile devices. Most of them focused on energy consumption. \mbox{\cite{li_investigation_2014}} directed a research to improve code practices and provided developers with guidelines to extend battery of mobile devices. \mbox{\cite{linares-vasquez_mining_2014}} mined Android energy-greedy
API usage patterns, making developers aware of
which components and APIs are more energy efficient. Similarly,
\mbox{\cite{manotas_seeds:_2014}} developed a set of recommendations to support developers in coding more
energy aware apps. \mbox{\cite{wan_detecting_2015}} presented a technique to detect display energy hotspots of mobile device apps, reporting them to developers.}

\comment{In addition to energy consumption, other performance metrics have also been studied. \mbox{\cite{gui_truth_2015}} quantified the impact of ads on energy consumption, but also on CPU, memory, and network usages on mobile device apps.  \mbox{\cite{chen_android_2016}} conducted a comprehensive study on the impact of languages,
compilers, and implementations on the performance of Android apps in terms of energy consumption and CPU usage. 
\mbox{\cite{hasan_energy_2016}} created detailed profiles of the energy consumed by common operations done on Java list, map, and set implementations. They also explored the memory usage of list implementations.} 

\comment{Several works have also studied the impact of user's decisions on software performance. \mbox{\cite{Amsel:2010:GTT:1753846.1753981}} proposed a tool for estimating the energy consumption of currently installed software systems. This tool determines which software systems are the most efficient given the user's 
current computer configuration.  It presents this information to the user in the form of a chart comparing the energy consumption and the CPU usage of software systems in the same category. 
\mbox{\cite{Procaccianti2011}} collected and analyzed power consumption data of a desktop computer simulating common usage scenarios. Automated by a Graphical User Interface (GUI) testing tool, they collected power consumption data by means of a power meter to extract an energy profile for each independent scenario. They obtained that usage patterns impact consistently on the energy consumption of software. \mbox{\cite{zhang_impact_2014}} analyzed the energy consumption of different apps and they highlighted the perils that users face and the ultimate responsibility users have for the battery life of their devices. 
For example, they obtained that a command line music player uses more than six times less energy than a different music player with a GUI, or that web-based desktop apps tend to consume more energy than non-web based. \mbox{\cite{behrouz_ecodroid:_2015}} proposed \mbox{EcoDroid}, an approach that estimates the energy cost of Android apps in a category and ranks them accordingly to help users make informed decisions. \mbox{EcoDroid} uses a combination of dynamic and static analyses and test cases to execute apps and estimates their energy cost based on their API usage. These estimates take into account the energy cost of the paths executed by test cases to rank Android apps in terms of energy efficiency. Recently, \cite{saborido_optimizing_2016} presented ADAGO, a simple recommendation system to propose optimal sets of Android apps that minimize power and data usages while maximizing the apps rating.}

\comment{The closest works to our approach are EcoDroid and ADAGO. Both share with APOA the idea of supporting users. 
However, APOA tackles a different problem to \mbox{EcoDroid} and \mbox{ADAGO}, with different variables (categories of apps in a marketplace) and objectives (metrics).}

\section{Conclusion}
\label{sec:conclusion}

The official marketplaces of Android and Apple platforms offer more than a million mobile device apps belonging to different categories. In both marketplaces information about performance metrics is not usually available and, therefore, mobile users select apps based on other criteria, as the rating. 
Even if performance metrics are available, different apps have different performance and, depending on the context of use, some metrics could be more important than others. For instance, users could prefer to optimize battery life and network usage in a foreign trip context while CPU, memory, and network usages could be preferred if the phone is plugged to the car charger in a navigation context. 
All of this makes the comparison of apps difficult in terms of performance because the required cognitive effort. APOA is proposed as a recommendation tool to complement mobile device app marketplaces allowing users and developers to compare optimal apps or to rank them relevant to their current context and needs. \comment{APOA takes as input metric values of apps and, therefore, it is totally independent to the measurement process.}  

We evaluated APOA over an Android case study. Out seven categories and 140 apps, we defined typical usage scenarios and we collected information about performance metrics.  We used APOA to show that selecting apps based on their rating is not an optimal choice in terms of performance. APOA was able to find several sets of Pareto optimal apps where eight of them had a similar rating but they improved performance metrics. The improvement for power, CPU, memory, and network usages was up to 3.66\%, 37.09\%, 12.33\%, and 67.57\%, respectively. In addition we also showed that, sacrificing the rating in a small magnitude, performance metrics are improved in a big proportion. We obtained that, by only sacrificing the rating by 7.16\% the improvement in performance metrics raised in 15.98\%, 67.99\%, 26.40\%, and 75.73\%,
respectively. We also showed the benefit of using APOA in different contexts of use to find optimal combinations of apps and compare the existing trade-off between performance metrics. Finally, we illustrated how the availability of performance metrics can be helpful for developers before they release a new app.\comment{ For example, comparing the performance of a new app with respect to its competitors.} This fact would motivate the development of more efficient apps, which benefits final mobile device app users.

\comment{Once our approach is implemented in a marketplace, it can be used on-demand by users. However, APOA can also be automatically used by marketplaces. This is feasible given that mobile device users usually link their devices to the app marketplace, such as the Google Play Store or the Apple Store apps. Thus, marketplaces know which apps the user has installed and APOA could use this information to make customized optimal recommendations for each user.}

As future work, we plan to apply the presented approach dynamically whereby user profiles of app use and their intent are mapped to the contexts that best fit them. Thus, app marketplaces would use user profiles and performance behavior to
recommend to users which apps they could download and consume. 
In addition to this, mobile device app developers could use inter-app comparison and performance measures to
continuously evaluate their apps performance in the app store market. When integrated into continuous integration, the developers could get
relative app ranking per each software change. By integrating comparison
into continuous integration (continuous inspection) developers could
maintain constant awareness of performance relevant issues their apps
might face.



\section*{Acknowledgements}
The authors would like to thank the Electrical Engineering department of Polytechnique Montreal for sharing their resources. In particular, we would like to thank Bryan Tremblay for his valuable help and support.

This work has been partially funded by the Spanish MINECO and
FEDER project TIN2014-57341-R (http://moveon.lcc.uma.es).
\balance
\bibliographystyle{elsarticle-harv}
\bibliography{usage,optimization,energy,rating,metaheuristics,urls}

\clearpage
\section*{Vitae}
\label{sec:vitae}
\textbf{Rub\'{e}n Saborido} received his BS. degree in Computer Engineering and his MS. in Software Engineering and Artificial Intelligence from University of Malaga (Spain), where he worked for three years as a researcher. In 2017, he received a Ph.D. in Computer Engineering from Polytechnique Montréal and his thesis was nominated for best thesis award. Rub\'{e}n research focuses on search based software engineering applied to performance and energy optimization of mobile devices. He is also interested in the use of metaheuristics to solve complex multiobjective optimization problems and in the design of algorithms to approximate a part of the whole Pareto optimal front taking into account user preferences. He has published seven papers in ISI indexed journals, and conference papers in MCDM, SANER, and ICPC. He co-organized the International Conference on Multiple Criteria Decision Making, in 2013.

\textbf{Foutse Khomh} is an associate professor at Polytechnique Montr\'{e}al, where he heads the SWAT Lab on software analytics and cloud engineering research (http://swat.polymtl.ca/). He received a Ph.D. in Software Engineering from the University of Montr\'{e}al with the award of excellence. His research interests include software maintenance and evolution, cloud engineering, empirical software engineering, and software analytic. He has published more than 100 papers in international conferences and journals, and his work has received one Most Influential Paper Award, three Best Paper Awards and multiple nominations for Best Paper Awards. He has served on the program committees of several international conferences and reviewed for top international journals such as EMSE, TSE and TOSEM. He is on the Review Board of EMSE. He is program chair for Satellite Events at SANER 2015, program co-chair of SCAM 2015 and ICSME 2018, and general chair of ICPC 2018. He is one of the organizers of the RELENG workshop series (http://releng.polymtl.ca) and has been guest editor for special issues in the IEEE Software magazine and JSEP.

\textbf{Abram Hindle} is an associate professor at the University of Alberta, in Edmonton, Alberta, Canada within the Department of Computing Sciences. He focuses his research on the evidence-based study of software development. His research combines the domains of performance analysis, energy consumption analysis (Green Mining), natural language processing, computer music, and information retrieval with software engineering. He has published several papers in international conferences and journals, including EMSE, ICSE, FSE, ICSM, MSR, and SANER. He has served on the program committees of several international conferences including ICSME, ICSE, MSR, and SCAM. 

\textbf{Enrique Alba} is a professor of computer science at the University of Málaga, Spain, where he leads the NEO (Networking and Emerging Optimization) group. His current research interests involve the design and application of evolutionary algorithms, ant colony optimization, particle swarm optimization, and other bio-inspired systems to real problems including telecommunications, software engineering, combinatorial optimization, and bioinformatics among others. Prof.  Alba  has  published  12  monographs  on  complex 
problem  solving,  more than 70  papers  in  ISI  indexed 
journals,  and  more  than  250  conference  papers.  He  has also  coordinated  several  national  and  international 
research projects in the past.  Finally,  Prof. Alba  works in the program committee
 of well-known  important conferences in several fields,  like  ACM  GECCO, EvoCOP, IEEE  CEC, IPDPS, PPSN, and  many  more,  as  well he has  organized 
international events like  GECCO 2013 (as general chair), IEEE/ACM MSWiM, and NIDISC. He  also  works  as  reviewer for  EJOR, Computer  Communications, IEEE  Transactions  (on  EC, Education, PDS, SMC), JMMA, Journal  of  Heuristics, JPDC,  PARCO,  etc.  Besides,  Prof.  Alba works in the editorial board of several international journals related to optimization, telecommunications, and 
parallel systems.

\end{document}